\documentclass[conference]{IEEEtran}
\usepackage{cite}

\ifCLASSINFOpdf
\usepackage[pdftex]{graphicx}
  \DeclareGraphicsExtensions{.pdf,.jpeg,.png}
\else
  \usepackage[dvips]{graphicx}
  \DeclareGraphicsExtensions{.eps}
\fi
\usepackage{amsmath}
\usepackage{algorithmic}

\usepackage{array}

\ifCLASSOPTIONcompsoc
  \usepackage[caption=false,font=normalsize,labelfont=sf,textfont=sf]{subfig}
\else
  \usepackage[caption=false,font=footnotesize]{subfig}
\fi
\usepackage{dblfloatfix}

\usepackage{url}

\newcommand{\ignore}[1]{}
\usepackage{multirow}
\usepackage{multicol}
\usepackage{pifont}
\usepackage{comment}
\usepackage{booktabs}
\usepackage{xspace}
\usepackage{enumitem}

\newcommand{\mpo}{$<_p$\xspace}
\newcommand{\mmo}{$<_m$\xspace}
\newcommand{\po}{<_p}
\newcommand{\mo}{<_m}

\newcommand{\mytodo}[1]{\todo[inline]{#1}}

\newcommand{\comms}[1]{\hl{#1}}
\newcommand{\cmark}{\ding{51}}
\newcommand{\xmark}{\ding{55}}

\newcommand{\vr}{\textit{vr}\xspace}
\newcommand{\lfvr}{\textit{lfvr}\xspace}
\newcommand{\vsb}{\textit{$v_{min, sb}$}\xspace}
\newcommand{\vlsq}{\textit{$v_{min, lsq}$}\xspace}
\newcommand{\orq}{\textit{orq}\xspace}

\newcommand{\bonit}[1]{\textit{#1}}
\newcommand{\method}{\textsc{Louvre}\xspace}

\usepackage{tikz}
\newcommand{\mynode}[1]
{\tikz\node[circle,scale=0.5,color=white,fill=black]{\large #1};}

\newtheorem{verule}{Versioning Rule}

\begin{document}
%
\title{Louvre: Lightweight Ordering Using Versioning for Release Consistency}

\author{
\IEEEauthorblockN{Pranith Kumar, Prasun Gera, Hyojong Kim, Hyesoon Kim}
\IEEEauthorblockA{{\it Georgia Institute of Technology}\\
{\it School of Computer Science, College of Computing}\\
{\it Atlanta, GA, USA}\\
{\it \{pranith, prasun.gera, hyojong.kim, hyesoon.kim\}@gatech.edu}}}

\maketitle
\begin{abstract} 
  Fence instructions are fundamental primitives that ensure consistency in a
  weakly consistent shared memory multi-core processor. The execution cost of
  these instructions is significant and adds a non-trivial overhead to
  parallel programs. In a na\'{i}ve architecture implementation, we track the
  ordering constraints imposed by a fence by its entry in the reorder buffer
  and its execution overhead entails stalling the processor's pipeline until
  the store buffer is drained and also conservatively invalidating speculative
  loads. These actions create a cascading effect of increased overhead on the
  execution of the following instructions in the program. We find these
  actions to be overly restrictive and that they can be further relaxed
  thereby allowing aggressive optimizations.

  The current work proposes a lightweight mechanism in which we assign
  ordering tags, called versions, to load and store instructions when they
  reside in the load/store queues and the write buffer. The version assigned
  to a memory access allows us to fully exploit the relaxation allowed by the
  weak consistency model and restricts its execution in such a way that the
  ordering constraints by the model are satisfied. We utilize the information
  captured through the assigned versions to reduce stalls caused by waiting
  for the store buffer to drain and to avoid unnecessary squashing of
  speculative loads, thereby minimizing the re-execution penalty. This method
  is particularly effective for the release consistency model that employs
  uni-directional fence instructions. We show that this mechanism reduces the
  ordering instruction latency by 39.6\% and improves program performance by
  11\% on average over the baseline implementation.

\end{abstract}


\ifCLASSOPTIONpeerreview
\begin{center} \bfseries EDICS Category: 3-BBND \end{center}
\fi
%
\IEEEpeerreviewmaketitle

\section{Introduction}

As core counts have continued to increase over the last decade, the shared
memory programming model has become an attractive choice for high performance,
scientific, and general purpose parallel applications. The use of this model
is facilitated by high-level libraries and frameworks such as
Cilk~\cite{blu:joe96} and OpenMP~\cite{dor:rod05}. The shared memory
programming model necessitates a consistent view of the memory for programmers
to reason about the accuracy of parallel programs. This consistent view is
dictated by the memory consistency model of the processor.

\ignore{ The memory consistency model is a contract between the processor and
  any program that runs on it and describes the types of memory re-orderings
  possible on the processor.}

Higher level languages standardize on a consistency model to maintain
portability across architectures. Languages such as C and C++ have
standardized on data-race-free memory models~\cite{adve:hill90} that guarantee
that accesses to synchronization objects (or \textit{atomics}) are
sequentially consistent. Sequential consistency for such accesses is ensured
by the compiler, which generates appropriate ordering instructions. These
instructions indicate an ordering point in the instruction stream, and when it
is encountered by the processor, certain classes of optimizations on the
following memory accesses are disabled to ensure
consistency~\cite{sev11,eid:reg08,tra:pra06}. Optimizations such as
speculative execution, prefetch, and buffering of store instructions are
implemented by modern processors to reduce the cost of memory accesses
~\cite{cho:fah04,leb:kop02,luk01}. The fence instructions restrict
optimizations which exploit instruction-level parallelism, disable speculative
execution of post-fence loads and stores, and introduce stalls to drain the
store buffer.

Researchers have proposed techniques that allow a processor to employ hardware
optimizations when ordering instructions are scheduled, thereby reducing their
execution overhead~\cite{mar:tor02,ran:pai97}. One such technique utilizes
check-pointing along with aggressive speculative execution of post-fence loads
and stores. The check-pointed state is tracked and updated so that it enables
roll back in case of ordering
violations~\cite{gni:fal99,dua:muz13,dua:fen09}. Another promising approach is
to increase the granularity of enforcing consistency, thereby amortizing the
overhead of ordering instructions. This technique has been applied both in
hardware~\cite{cez:tuc07} and at an algorithmic
level~\cite{lee:sim15,vor:kod14}. Further, to increase the scope for
speculative execution based on the observation that ordering violations rarely
occur at runtime~\cite{gni:fal99}, studies have proposed techniques that
identify scenarios in which violations are likely to occur and handle them in
special ways. These techniques use annotations that tag ordering
instructions~\cite{lin:nag13}, hardware extensions~\cite{dua:muz13}, or
run-time information~\cite{dua:hon15}. However, all these techniques were
proposed in the context of stronger memory consistency models such as that of
x86 and little work has explored reducing the overhead of ordering
instructions, specifically in weak memory model architectures.

Release consistency ($RC_{sc}$)~\cite{gha:len90} is one of the most widely
implemented weak memory model in recent processors. Variants of this model
have been adopted in architectures such as Itanium, ARM64, and PowerPC and it
is also being considered for adoption in RISC-V
processors~\cite{wat:lee16}. In this memory model memory accesses are
classified as ordinary, acquire, or release operations. This classification is
done using special instructions in the ISA.  A \textit{load-acquire} operation
and a \textit{store-release} operation place ordering constraints either on
the memory accesses following it or preceding it respectively.  This is in
contrast to a full fence which places ordering constraints on both the memory
accesses. These acquire and release operations allow more leeway for possible
re-ordering of the memory accesses, thereby allowing a processor
implementation to reduce the overhead of ensuring memory consistency.

We propose \method, a low-overhead hardware extension that ensures memory
ordering using {\em versioning} and helps in reducing the overhead of fence
instructions in a processor implementing the release consistency model. Our
proposed mechanism adds ordering tag fields to the load/store queues and the
write buffer. In the proposed technique, we assign versions to memory
instructions when they are issued. We update the version being assigned upon
issuing an ordering instruction. Instead of tracking the ordering constraints
by the entry of the fence instruction in the reorder buffer (ROB), we track
them in a separate FIFO queue. This allows us to retire the ordering
instruction from the ROB while preserving the constraint that needs to be
enforced. Using the version assigned to memory accesses, we identify and
rectify any consistency violations at run-time without the need for
maintaining any global state or inter-core communication. Compared to
versioning, traditional implementation of ordering instructions stall the
pipeline to drain the store buffer to ensure order among stores in the buffer
and memory accesses following the ordering instruction. In the proposed
versioning mechanism, we issue and execute instructions that follow the
ordering instruction without draining the store buffer. Since we version the
stores in the store buffer, we enforce the correct ordering among them
utilizing these versions. Furthermore, the versioning mechanism is efficient
for one-way fence instructions, since it allows the processor to speculatively
issue and execute instructions following a store-release instruction before
the store-release instruction itself has finished execution.

Although we did not find any hardware optimizations in the literature
specifically targeting one-way fence instructions, the work which is closest
in spirit to our current work is zFence~\cite{aga:sin15}. As in the current
work, the authors of the zFence paper focus on reducing the stalls caused by
waiting for the store buffer to drain. They achieve this by introducing a new
coherence state which indicates exclusive permission to a cache line without
the cache line being available. When all the stores waiting in the store
buffer acquire this exclusive permission, you do not need to drain the store
buffer since no other processor can modify those cache lines, thereby reducing
the fence overhead. In contrast, the current proposal requires only minor
hardware changes to avoid the store buffer drain stall.

The main contributions of our work are as follows:

\begin{itemize}
\item We describe a detailed micro-architectural description for implementing
  $RC_{SC}$ memory model semantics.
\item We propose a hardware mechanism based on versioning loads and stores
  that reduces the cost of ordering instructions in weak memory model
  architectures.
\item We propose a low overhead mechanism that optimizes the execution of
  uni-directional ordering instructions by taking advantage of re-orderings
  allowed in a weak memory model architecture.

\end{itemize}


\section{Background and Motivation}

\subsection{Fence instructions}

\ignore{ Ordering instructions on x86 and ARM64 have different consistency
  semantics. \bonit{Explicit} ordering instructions such as \textit{mfence} in
  x86 or \textit {dmb} in ARM64 ensure that memory accesses before the fence
  in program order are architecturally visible~\cite{sorin2011primer} before
  any memory access after the fence. Their only purpose is to ensure a
  specific ordering among memory accesses. \bonit{Implicit} fence instructions
  enforce consistency along with instruction semantics. An example of an
  implicit ordering instruction is the \textit{xchg} instruction in x86.  This
  instruction is used to exchange values between two locations and also has
  the semantics of an \textit{mfence} instruction. We can also add a
  \textit{lock} prefix to certain x86 instructions so that they implicitly
  enforce ordering~\cite{IA32}. Because of the implicit ordering semantics of
  \textit{xchg} and \textit{lock} prefixed instructions, they form the basis
  of implementations of synchronization on x86 processors. The classification
  of fences for two popular ISAs, x86 and ARM64, is listed in
  Table~\ref{tab:fence_classification}.  }

Ordering instructions can be classified as either \textit{uni-directional}
(also known as \textit{one-way} fences) or \bonit{bi-directional} (also known
as \textit{full fences}). Instructions \textit{mfence} and \textit{dmb} are
bi-directional in that they do not allow any re-ordering across them. The
restrictions placed by such fence instructions on memory accesses conflict
with the optimization techniques used to hide the latency of long memory
accesses in modern processors. Efforts to reduce this overhead have led to the
design of the release consistency model~\cite{gha:len90}. This model
prescribes acquire and release semantics, which are achieved by
uni-directional fence instructions. These instructions allow re-ordering of
memory accesses across them in only one direction and have either acquire or
release semantics. Slight variations of this conventional consistency model
are implemented in architectures such as Itanium and ARM64. On ARM64, a
\textit{load-acquire} instruction (also called as a synchronizing load,
\textit{ldar}) enforces an order between its load and all the following memory
accesses. A \textit{store-release} instruction (also called as a synchronizing
store, \textit{stlr}) enforces an order between its store and all the
preceding memory accesses. A store-release also ensures that the store is
multi-copy atomic; a store is made visible to all other processors
simultaneously~\cite{flu:gra16}. Additionally, a fence instruction is
sequentially consistent with respect to all other fence instructions.

\begin{figure}
\includegraphics[width=0.5\textwidth]{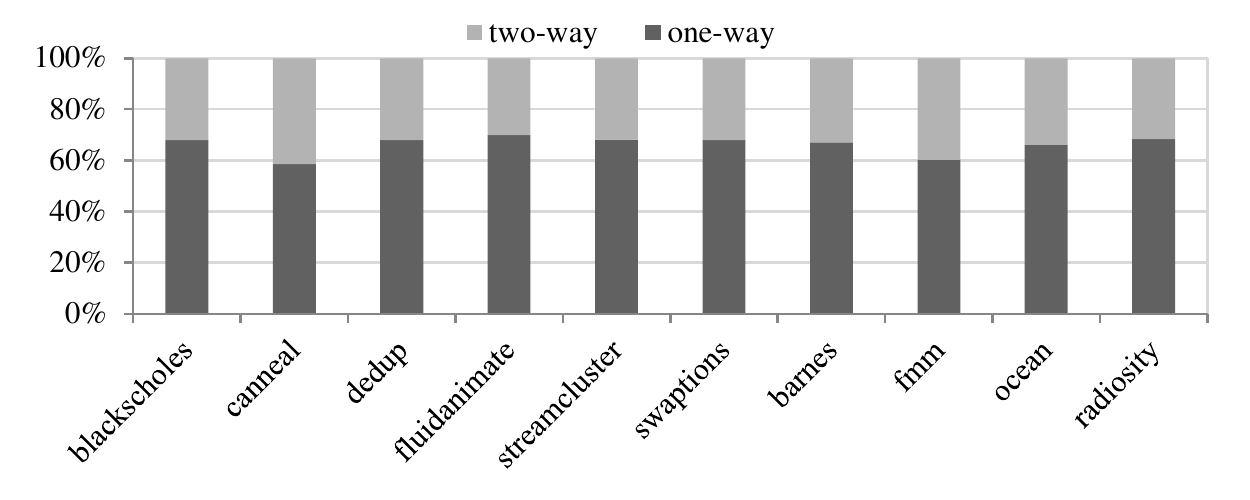}
\caption{Distribution of fences in ARM64}
\label{fig:fence_ratio}
\end{figure}

\ignore{
\begin{table}
  \caption{Fence instructions in x86 and ARM64 ISA}
  \begin{center}
  \begin{tabular}[width=0.5\textwidth]{clll}
    \toprule
    ISA & Instruction & Constraints & Type \\
    \midrule
    ARM64 & \begin{tabular}{@{}l}ldar \\ stlr\\dmb \end{tabular}
          & \begin{tabular}{@{}l} One-way \\ One-way \\ Two-way \end{tabular}
          & \begin{tabular}{@{}l}Implicit \\ Implicit \\ Explicit \end{tabular} \\
    \midrule
    x86 & \begin{tabular}{@{}l}mfence\\ xchg \\ lock(prefix)\end{tabular}
        & \begin{tabular}{@{}l}Two-way\\Two-way\\Two-way \end{tabular}
        & \begin{tabular}{@{}l}Explicit\\Implicit\\Implicit \end{tabular} \\
    \bottomrule
  \end{tabular}
  \end{center}
  \label{tab:fence_classification}
\end{table}
}

Since one-way fences remove constraints on possible memory access re-ordering,
generating them is more efficient than generating full fences whenever
possible.  Figure~\ref{fig:fence_ratio} shows the frequency distribution of
one-way and full fences in benchmarks from the Parsec~\cite{parsec} benchmark
suite compiled for ARM64 processors.  We observe that one-way fences, on
average, are twice as frequent as full fences in most of the benchmarks.
Exploiting reduced constraints enforced by a one-way fence necessitates an
efficient micro-architectural implementation.  However, optimizations specific
to one-way fences are limited.  As a result, a uni-directional fence, when
treated as a full fence limits the potential performance improvement. To the
best of our knowledge, \textit{our proposal is the first to present a detailed
  optimal micro-architectural implementation of one-way fences}.

\subsection{Definitions}
\label{sec:def}

\begin{figure}[h]
\includegraphics[width=0.5\textwidth]{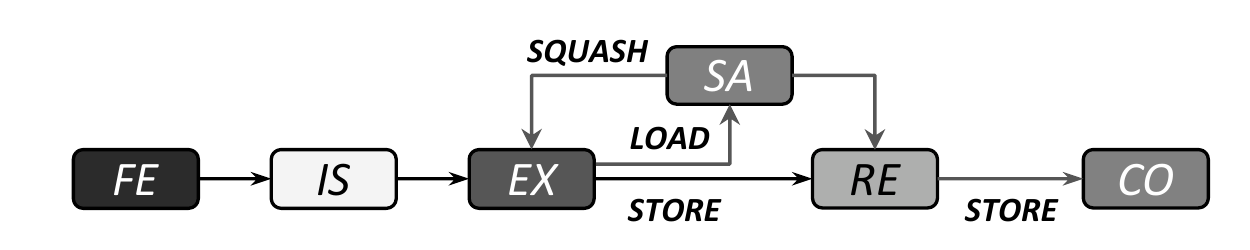}
\caption{Life-cycle of a memory access instruction.} 
\label{fig:lifecycle}
\end{figure}

In the rest of the paper, we use the following terminology. An instruction is
fetched (FE), decoded, and then issued (IS). Once it is issued (possibly
out-of-order), it can be executed (EX) when all its dependencies are met. Once
an instruction reaches the head of the reorder buffer (ROB) and finishes
execution, it can \textit{commit or retire} (RE). Upon retiring, a store
instruction moves from the ROB to the store buffer (SB), at which point it is
said to be a \textit{post-commit} store. A store will \textit{complete} (CO)
once it updates the value in the cache and drains from the store buffer.  A
load is said to be \textit{satisfied} (SA) after it loads the value into a
register. While a load can retire only after it is satisfied, a store can
retire before it completes. An instruction remains in a \textit{speculative
  state} as long as it resides in the ROB. The life cycle of a memory access
instruction is shown in Figure~\ref{fig:lifecycle}. Please note that a
satisfied load in a speculative state can be squashed (SQ) and re-executed
along with all of its dependent instructions.

Program order (po) is the order in which instructions are fetched and decoded
by the same core. We use \textit{po-before/previous} to refer to instructions
that are earlier in program order and \textit{po-after/later/following} to
refer to instructions that are later in program order relative to the current
instruction.

\subsection{RC\textsubscript{SC} Semantic Rules}
\label{sec:rc_rules}

The following semantic rules describe the constraints imposed by
$RC_{SC}$~\cite{gha:len90}.  Here $X$ and $Y$ are two instructions, \mpo and
\mmo are the program and global memory order respectively. $L(a)$ and $S(a)$
are load and store to address $a$ respectively. $FF(X)$ is true if $X$ is a
full fence, $LDAR(X)$ is true if $X$ is a load-acquire fence, $STRL(X)$ is
true if $X$ is a store-release fence, and $LOAD(X)$ or $STORE(X)$ is true if
$X$ is a load or store instruction respectively.

\begin{enumerate}[label=RC\arabic*.,leftmargin=1.5cm]
\item $ X \po Y,\; FF(X) \implies X \mo Y $
\item $ X \po Y,\; FF(Y) \implies X \mo Y $
\item $ X \po Y,\; LDAR(X) \implies X \mo Y $
\item $ X \po Y,\; STRL(Y) \implies X \mo Y $
\item $ X \po Y,\; X, Y \in (FF,\; LDAR,\; STRL) \implies X \mo Y $
\item $ Value\; of\; L(a) = Value\; of\; Max_{<m}\; \{S(a)\; |\\
  S(a) \po L(a)\; or\; S(a) \mo L(a)\} $
\end{enumerate}

In $RC_{SC}$, unlike in sequential consistency ($SC$), program order does not
imply global order for ordinary memory accesses. A synchronizing load
(\textit{load-acquire}), a synchronizing store (\textit{store-release}), and a
fence instruction enforce order among the memory accesses.

$RC1$ and $RC2$ specify that memory accesses that are po-before a full fence are
ordered before the po-later memory accesses.

$RC3$ specifies that the load of a load-acquire fence should be ordered before
any po-later memory accesses.

$RC4$ specifies that the store of a store-release fence should be ordered
after all the po-previous memory accesses.

$RC5$ specifies that all the fence instructions should be ordered in sequential
order.

Finally, $RC6$ specifies that the value of a load is satisfied from the latest
po-before store to the same address pending in the store buffer. If there is
no such store, the load is satisfied from the latest global store to the same
address.

We later show in Section~\ref{sec:versioning_semantics} that \method satisfies
the $RC_{SC}$ semantics listed here.

\begin{table}[b]
  \caption{Constraints enforced by different ordering instructions in a
    conventional Release Consistency Architecture}
  \begin{tabular}{ c | c|c}
    \toprule
    Retirement of fence & Store buffer & On cache invalidation \\
    \midrule
    load-acquire & $-$ & Squash load \\
    store-release & Drain & $-$ \\
    full & Drain & Squash load \\
    \bottomrule
  \end{tabular}
  \label{tab:overhead}
\end{table}

\subsection{Conventional Microarchitecture}

\begin{figure*}[t!]
\includegraphics[width=\textwidth]{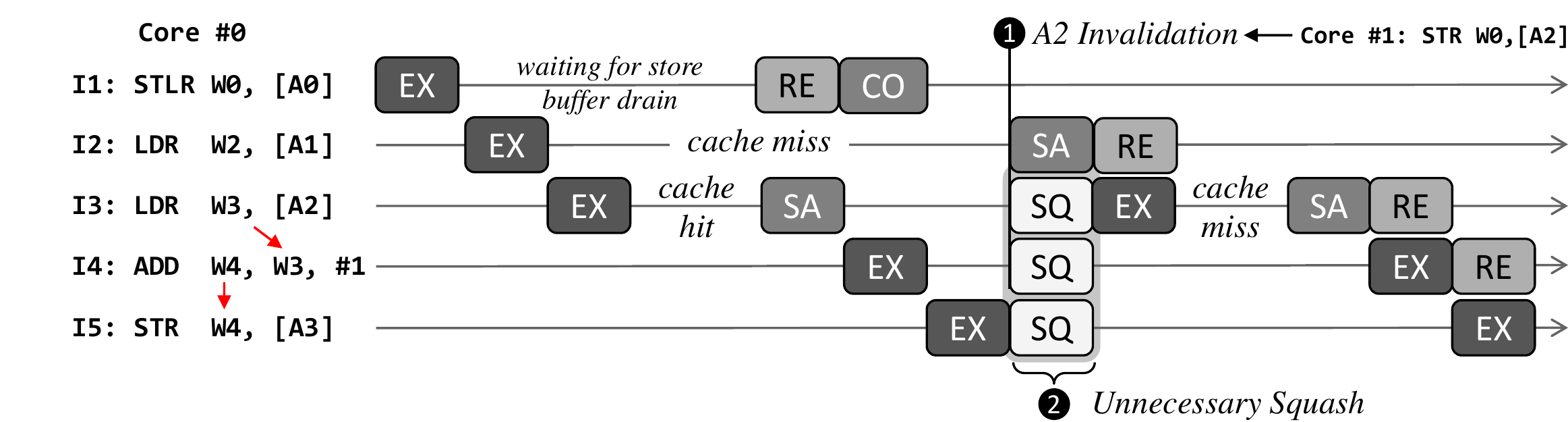}
\caption{Inefficient load speculation in a weak consistency architecture.}
\label{fig:load_speculation}
\end{figure*}

In this section, we describe three actions that a conventional processor
performs to avoid ordering violations and to implement the semantics of
$RC_{SC}$ model listed in the previous section: retirement of a fence,
draining the store buffer, and squashing a speculative load on cache line
invalidation. The processor enforces $RC1$ for a full fence by draining the
store buffer before retiring the full fence instruction. This ensures that
post-fence stores do not complete before pre-fence stores from the unordered
store buffer. The processor enforces $RC2$ by squashing any speculative
post-fence load upon receiving an invalidation request to the cache line until
the full fence is retired. The one-way fences act as two individual halves of
the full fence. The processor performs the same actions separately to enforce
rules $RC3$, $RC4$, and $RC5$. To enforce $RC3$ for a synchronizing load, the
processor needs to squash any speculative po-later loads upon invalidation as
long as the synchronizing load is in-flight. To enforce $RC4$ for a
synchronizing store, the processor needs to drain the store buffer before
retiring the synchronizing store instruction. This prevents the synchronizing
store from completing ahead of the stores in the store buffer. These actions
also ensure that the rule $RC5$ is followed. Finally, for rule $RC6$, the
value of a load is satisfied from the store buffer if a store to the same
address is pending in it; otherwise, a request is sent to the memory
hierarchy. We summarize the constraints that need to be satisfied for each
type of ordering instruction to be retired in Table~\ref{tab:overhead} .

\section{Motivation}
\label{sec:inefficiency}

In this section we discuss the ordering constraints imposed by different fence
instructions and identify two major sources of overhead caused by these
constraints.

\subsection{Constraints of ordering instructions}
\label{sec:overhead}

A full fence and a load-acquire fence enforce the constraint that the
processor squash all speculative loads satisfied from a cache line when that
cache line is invalidated. This constraint prevents speculative loads from
using stale values, which would otherwise not be used if the loads were
executed non-speculatively. A full fence and a store-release fence enforce the
constraint that the processor drain the store buffer of all po-previous stores
before retiring the fence. In the case of a full fence, this constraint
prevents a po-later store from completing ahead of a po-previous store. In the
case of a store-release fence, this constraint prevents the synchronizing
store from completing ahead of any po-previous stores. The inefficiencies caused
by these constraints are discussed in the following sections.

\ignore{
\begin{figure*}[t!]
\includegraphics[width=0.5\textwidth]{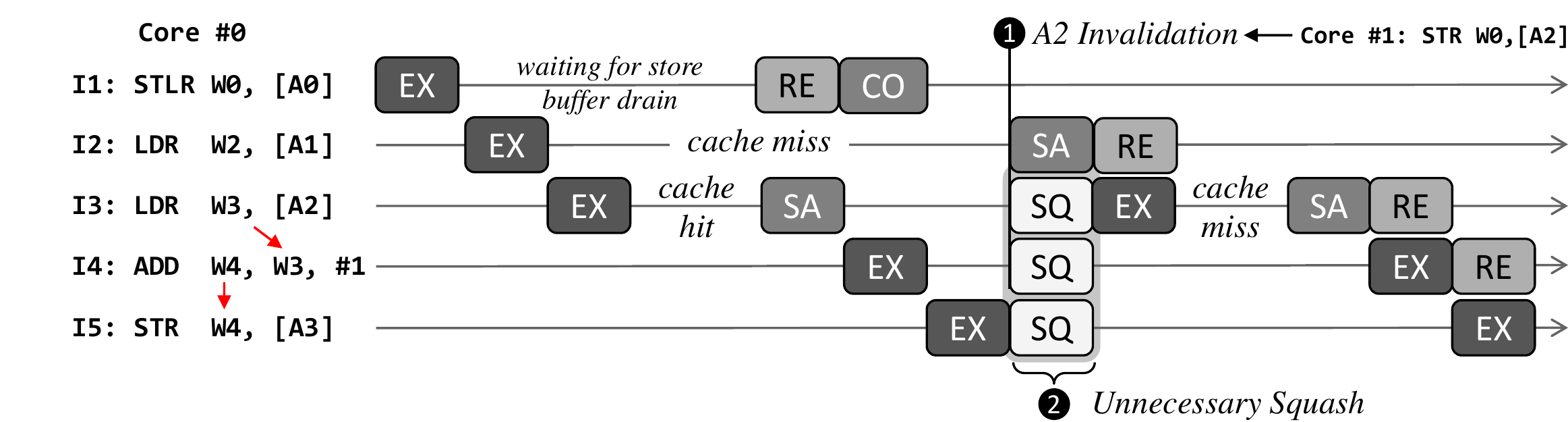}
\caption{Inefficient load speculation in a weak consistency architecture.}
\label{fig:load_speculation}
\end{figure*}
}

\subsection{Inefficient load speculation}
\label{sec:load_spec}

A load-store queue (LSQ) keeps track of all in-flight load and store
instructions in program order. To detect ordering violations, the LSQ snoops
the addresses of the incoming cache invalidations. If the address of a cache
invalidation matches that of an in-flight satisfied load, then the load might
have read a stale value causing an ordering violation. In such cases, the
processor squashes the load and all later instructions, and re-executes them.

\ignore{ An entry is allocated to each memory
  instruction in the LSQ at the time it is issued into the instruction window
  (ROB). The LSQ also forwards in-flight store values to matching loads and
  detects ordering violations. Upon execution of a store instruction, the LSQ
  snoops for a matching load that is later in program order but that was
  executed earlier. If the LSQ finds such a load, it squashes the load and all
  later instructions and flushes the pipeline. Similarly, upon execution of a
  load instruction, the LSQ is searched for a matching store that is earlier
  in both program and issue order and ensures that the load receives the value
  forwarded from the matching store.} 

The first inefficiency we identify is the unnecessary invalidation of
speculative loads to prevent ordering violations in a processor implementing
release consistency. While loads require squashing in stricter memory model
processors, it is not always necessary in weaker model processors. In
particular, the processor needs to squash a speculative load \textit{only} in
the presence of an in-flight fence enforcing order on the load. Since ordering
instructions retire after they enforce the required constraints, a speculative
load no longer violates ordering when all preceding fence instructions are
retired. In such cases we do not need to squash it on an invalidation. These
scenarios arise in the presence of all three kinds of fences. We detail one
such scenario with a synchronizing store next. An example with a
synchronizing load is illustrated in Section~\ref{sec:ver_example}.

In Figure~\ref{fig:load_speculation}, instruction $I1$ is a store-release
fence. Instructions $I2$ and $I3$ are independent loads that access addresses
$A1$ and $A2$ respectively. Instructions $I4$ and $I5$ are dependent on the
value loaded by instruction $I3$. The illustration shows a situation in which
$I2$ has a cache miss but $I3$ has a cache hit. Because of the cache hit, the
speculative load $I3$ is satisfied earlier than the speculative load $I2$.
The synchronizing store $I1$ is waiting for the store buffer to drain to
prevent possible reordering with any pending stores residing in the store
buffer. At this point, an invalidation request for cache line containing $A2$
arrives (say from core 1) at \mynode{1}. Note that this is after the
speculative load $I3$ is satisfied. However, since $I3$ is still speculative,
the processor squashes it and its dependent instructions, $I4$ and $I5$, and
re-executes them. However, since there is no fence enforcing order on $I3$,
the value loaded by it does not violate any ordering constraints. So squashing
and re-execution of instructions $I3$, $I4$, and $I5$ is not necessary.

\ignore{
A similar situation that illustrates unnecessary squashes in the presence of a
load-acquire fence is give in
Figure~\ref{fig:loadacquire_speculation_combined}.

\begin{figure}
\includegraphics[width=0.5\textwidth]{figures/loadacquire_speculation_combined}
\caption{Load speculation in presence of load-acquire fence}
\label{fig:loadacquire_speculation_combined}
\end{figure}
\begin{figure}
\includegraphics[width=0.5\textwidth]{figures/loadacquire_speculation}
\caption{Inefficient load speculation in presence of load-acquire fence}
\label{fig:loadacquire_speculation}
\end{figure}
\begin{figure}
\includegraphics[width=0.5\textwidth]{figures/loadacquire_speculation_versioning}
\caption{Efficient load speculation in presence of load-acquire fence}
\label{fig:loadacquire_speculation_versioning}
\end{figure}
}

We identify such situations in \method and avoid the squashing and
re-execution penalty. This is explained later in
Section~\ref{sec:spec_exec_inv}.

\subsection{Inefficient store retirement} 
\label{sec:store_ret}

\begin{figure*}
\includegraphics[width=\textwidth]{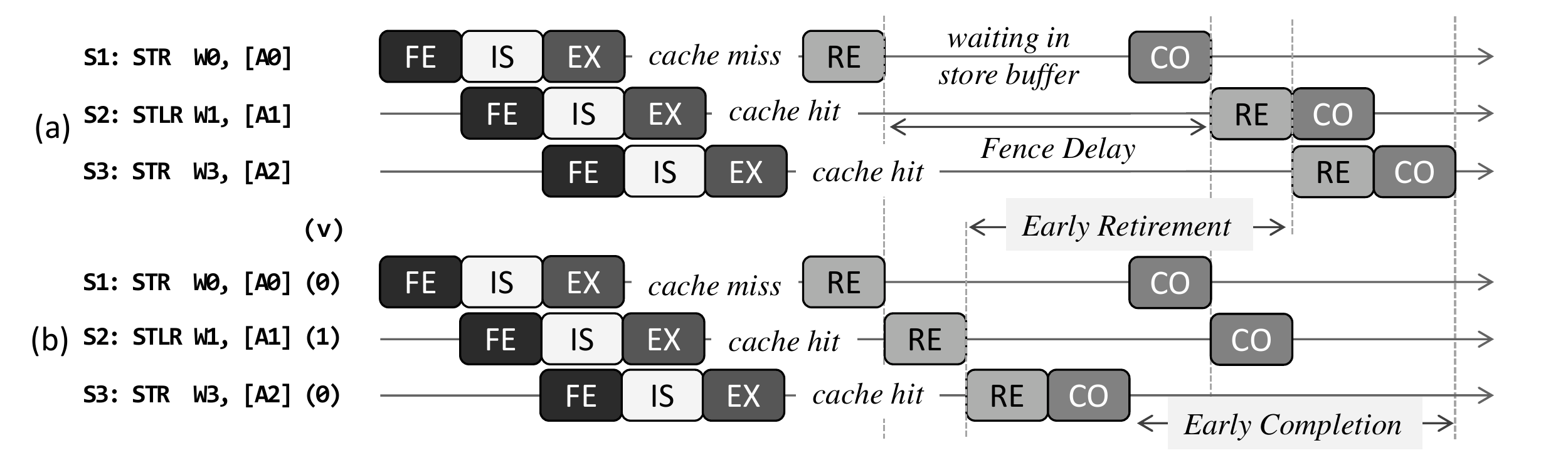}
\caption{Early retirement and completion of stores from an unordered store
  buffer (a) baseline and (b) optimal with versions (ver)}
\label{fig:early_store_retirement}
\end{figure*}

Typically, a store instruction that misses in the cache waits in the ROB until
the cache line is ready for an update. When this store reaches the head of the
ROB, instead of holding up the following instructions waiting for the cache
line, the processor retires it and buffers it in the store
buffer~\cite{cho:spr05, cho:fah04}. A store retires from the ROB and moves to
the store buffer from where it drains after completion~\cite{wen:ail07,
  bha:joh00}. In architectures implementing the TSO memory model, the store
buffer is ordered; the stores drain out from the buffer in FIFO order. In
weaker memory model architectures the store buffer can be unordered, that is,
the stores can drain out in any order. In our experiments, we assume a weak
memory model architecture and an unordered store buffer. Additionally, the
store buffer is augmented with a CAM structure that tracks the destination
address and relative age of the store in the buffer.  This is necessary for
store-to-load forwarding.

The second inefficiency we focus on is the stall of a full fence or
synchronizing store at the head of the ROB, while waiting for the stores in
the unordered store buffer to drain. This stall prevents ordering violations
in which post-fence stores are ordered before pre-fence stores. Without this
stall, a processor, after retiring the fence will retire post-fence stores and
place them in the store buffer along with pre-fence stores. Since the store
buffer is unordered, these post-fence stores can drain before pre-fence stores
resulting in an ordering violation.

The cost to drain the store buffer is the number of cycles it takes to service
all the cache misses of the pending stores in the buffer. Stalling execution
to drain the store buffer undermines the whole reason for including a store
buffer: to remove stores from the critical path in the processor
pipeline. Further, this stall adds to the latency of all instructions that
follow this fence instruction in the ROB. If the ROB is full, this delay may
even propagate to the instruction fetch stage.

A synchronizing store allows memory accesses following the instruction to be
ordered before it.  Figure~\ref{fig:early_store_retirement} illustrates an
example scenario in which $S1$, $S2$, and $S3$ are independent stores with
$S2$ being a synchronizing store. Figure~\ref{fig:early_store_retirement}(a)
shows how, in a conventional implementation, $S2$ blocks store $S3$ from
retiring from the ROB while waiting to drain the store buffer containing store
$S1$. Here store $S1$ has a cache miss, so the cost of draining the store
buffer is the latency of fetching this cache line. Once store $S1$ completes,
$S2$ and $S3$ can retire. Since no ordering constraint is imposed by the
synchronizing store $S2$ on $S3$, it can complete earlier. Using versioning,
we retire stores $S1$, $S2$, and $S3$ once they reach the head of ROB and
place them in the store buffer. The assigned version ensures that $S2$
completes only after $S1$ and that we place no constraint on $S3$. As shown in
Figure~\ref{fig:early_store_retirement}(b), versioning allows store $S3$ to
complete earlier than was possible in
Figure~\ref{fig:early_store_retirement}(a). Also, note that retiring $S2$ and
$S3$ without any removes the bottleneck on all following instructions.

\section{LOUVRE}
\label{sec:louvre_main}

In this section we describe the semantics of \method and explain how we employ
the versioning rules to order loads and stores.

\subsection{Semantic Rules}
\label{sec:versioning_semantics}

This section details the versioning semantic rules ($VSR$) that describe the
global order enforced by versions. Here $X$ and $Y$ are two instructions;
$v_x$ and $v_y$ are the versions assigned to those instructions. $LDAR(X)$ is
true if $X$ is a load-acquire instruction. $L(a)$ and $S(a)$ are load and
store to address a respectively.

\begin{enumerate}[label=VSR\arabic*.,leftmargin=1.5cm]
\item{$ X \po Y,\; v_x < v_y \implies X \mo Y $}
\item{$ X \po Y,\; LDAR(X)\; and\; v_x = v_y \implies X \mo Y $}
\item{$ Value\; of\; L(a) = Value\; of\; Max_{<m} \{S(a) | \\S(a) \po L(a)\; or\; S(a) \mo L(a)\}$}
\end{enumerate}

The crux of \method is the rule $VSR1$. Intuitively, it states that, if two
memory accesses that are in program order have increasing versions, then those
accesses should have the same global order.

\ignore{
Rule 2 is an artifact of the in-order retirement of instructions. It states
that if a store instruction is after a load instruction in program order, then
that store will be ordered after the load. This is trivially true because a
load becomes visible once it is retired, and a po-later store will retire and
be visible only after the load is visible.
}

$VSR2$ states that in versioning, if $X$ is po-before $Y$ and $X$ is a
load-acquire instruction, then it is ordered before $Y$, even though the
versions assigned to $X$ and $Y$ are the same. We explain this rule in detail
in Section~\ref{sec:ldar_fence_update}.

\ignore{
not assign increasing versions to po-later memory accesses for a load-acquire
fence to allow all possible reorders. We explain this versioning rule in
detail in Section~\ref{sec:ldaq_fence_update}. Briefly, if $X$ is a $LDAR$ and
$Y$ is a store, the processor ensures the required order in this case by the
order of retirement of $LDAR$ and the store $Y$. The store $Y$ will be visible
after retirement which is in-order after load-acquire $X$ retires and becomes
visible. In case $Y$ is a load and is satisfied ahead of $X$ becoming visible,
then squashing $Y$ on receiving an invalidation request and re-executing it
ensures the required ordering semantics of $RC_{SC}$.
}

VSR3 states that the value of a load is satisfied from the latest po-before
store to the same address pending in the store buffer. Otherwise the load is
satisfied by the latest store to that address from the memory system.

It is straightforward to see that versioning semantic rules are equivalent to
the semantic rules of $RC_{SC}$ listed in Section~\ref{sec:rc_rules}. $VSR1$
satisfies $RC_{SC}$ rules $RC1$, $RC2$, $RC4$, and $RC5$. $VSR2$ satisfies
$RC3$ whereas $VSR3$ is the same as RC6.

\subsection{Versioning Rules}
\label{sec:vr_update}

We explain the rules used to assign versions to memory accesses in this
section. These rules describe how the versions assigned to memory accesses
differ based on the ordering that needs to be enforced. Given two regular
memory accesses $a$ and $b$ that are issued from the same processor, $v_a$ and
$v_b$ are the versions assigned to them. Program order (po) is the order in
which instructions are issued by the processor.  $a\: {\tiny \po}\: b$ implies
that a is before b in program order.

\subsubsection{Full Fence}
\label{sec:ff_update}

A full fence should prevent any reordering across the fence (RC1 and RC2
listed in Section~\ref{sec:rc_rules}). When we issue a full fence instruction,
we assign versions such that all po-later memory accesses get a higher version
than the version assigned to all the po-previous memory accesses. This ensures
that we order all post-fence memory accesses after the pre-fence memory
accesses.

\begin{verule}
\label{lem:fence_version}
If a full fence separates two regular memory accesses a and b, then the
versions assigned for the memory accesses increases as follows:
\begin{center}
$  a \po fence \po b \implies v_a < v_b$
\end{center}
\end{verule}

\subsubsection{Synchronizing Load}
\label{sec:ldar_fence_update}

\begin{figure}[!h]
\includegraphics[width=0.5\textwidth]{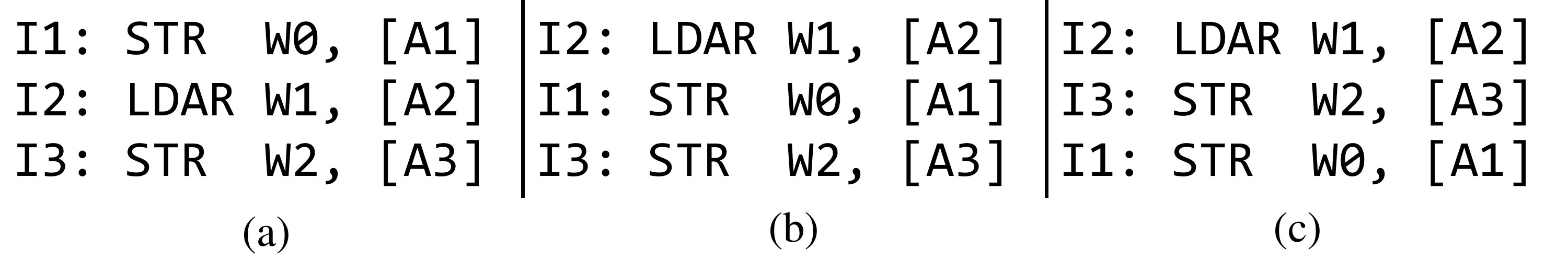}
\caption{The valid reorders for instructions in program order in (a) are (b)
  and (c).}
\label{fig:inefficient_ldaq_versions}
\end{figure}

A synchronizing load prevents any po-later memory accesses from being ordered
before it (Rule $RC3$ in Section~\ref{sec:rc_rules}). It differs from a full
fence in that the processor enforces the ordering constraint only between the
synchronizing load and all po-later memory accesses and not on po-previous
memory accesses. However, if we increase the version assigned to po-later
memory accesses like we do in the case of a full fence, we will impose
unnecessary ordering constraints on these accesses. The ordering constraint of
this fence should be valid \textit{only} as long as the synchronizing load is
speculative. Once the synchronizing load retires from the ROB, the
corresponding ordering constraint is satisfied. No further restrictions should
exist between the po-later memory accesses and the po-previous memory
accesses, i.e., stores residing in the store buffer. We illustrate an example
of such a situation in Figure~\ref{fig:inefficient_ldaq_versions}. In this
example, if we assign a higher version to store $I3$ than the version assigned
to store $I1$, we prohibit a valid reordering of $I1$ and $I3$ as listed in
Figure~\ref{fig:inefficient_ldaq_versions} (c). So as to not place such
unnecessary constraints, we do not increase the version assigned to
$I3$. Instead, we rely on the retirement order of instructions and speculative
load squashing in versioning to ensure the required memory order.

Consider $X <_p Y$ where $X$ is a load-acquire instruction. If $Y$ is a store
instruction, it will naturally be ordered after $X$. This is because a store
will be globally visible, and hence ordered, only after it retires that is in
turn after the retirement of the instruction $X$. Now if $Y$ is a load
instruction that is speculatively satisfied before $X$ retires, we squash $Y$
and re-execute it upon receiving a cache invalidation. This re-execution will
ensure that $Y$ is ordered after $X$.

\ignore{

In versioning, this ordering constraint is \textit{implicitly} satisfied by
the restriction on retirement of a load-acquire instruction.  Loads that are
after the load-acquire fence in program order and that are speculatively
executed will be squashed upon receiving a cache invalidation preventing them
from using stale values causing ordering violations. Stores that are after the
load-acquire fence in program order will retire in-order after the fence has
retired and hence will be visible only after the load associated with acquire
fence instruction is visible. Placing post-fence stores along with pre-fence
stores in the store buffer, after retiring the load-acquire fence, does not
violate ordering enforced by the fence because no ordering constraint exists
between these stores after load-acquire fence retirement.

In this way, the constraint on retirement ensures proper ordering of both
post-fence loads and stores with the load-acquire operation. Therefore, upon
issuing a load-fence instruction, we do not update the \vr register, thereby
assigning the same version to later memory accesses.

}

\begin{verule}
\label{lem:ldaq_version}
If a load-acquire fence separates two regular memory accesses a and b, then
the versions assigned for all the three memory accesses are the same.
\begin{center}
$  a \po ldar \po b \implies v_a = v_{ldar} = v_b$
\end{center}
\end{verule}

\subsubsection{Synchronizing Store}
\label{sec:stlr_fence_update}

\begin{figure}[!h]
\includegraphics[width=0.5\textwidth]{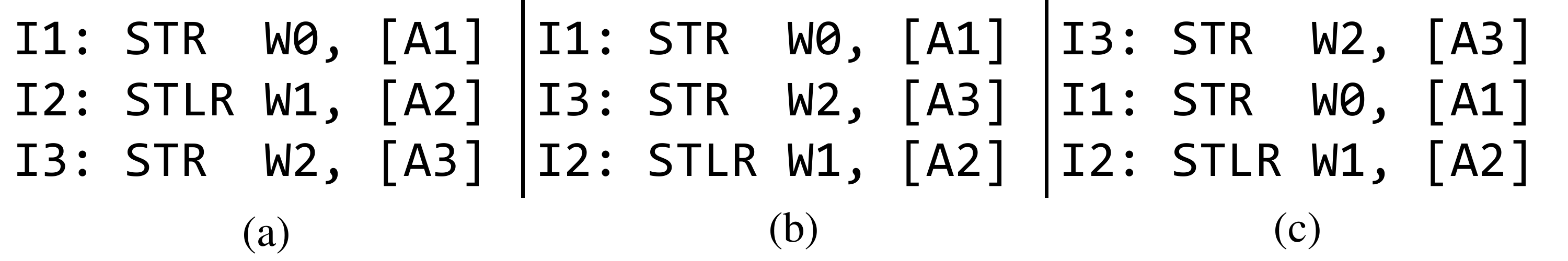}
\caption{The valid reorders for the instructions in program order in (a) are
  (b) and (c).}
\label{fig:inefficient_stlr_versions}
\end{figure}

A synchronizing store needs to be ordered after all the po-previous memory
accesses (Rule $RC4$ in Section~\ref{sec:rc_rules}). It places no constraint
on the po-later memory accesses. Our reasoning for not increasing the version
on issuing a store-release fence is similar to the one presented for the
synchronizing load. If we increase the version assigned to po-later memory
accesses than the synchronizing store, we place an unnecessary constraint on
them that should not remain once the synchronizing store is complete. We
illustrate an example of such a situation in
Figure~\ref{fig:inefficient_stlr_versions}. In this example, if we assign a
higher version to store $I3$ that follows the store-release fence than the
version assigned to store $I1$, we prohibit the valid reordering of stores
$I1$ and $I3$ listed in Figure~\ref{fig:inefficient_stlr_versions} (c).

To ensure that we place the required minimal ordering constraint, we assign a
higher version only to the synchronizing store. All po-later memory accesses
continue receiving the same version as the po-previous memory accesses. This
assignment ensures that we place the constraint only on the synchronizing
store.  Since all the post-fence stores receive the same version as pre-fence
stores, no constraint in placed on them and they can complete in any order
from the store buffer.

\ignore{
Hence, we update only the \lfvr register upon issuing a store-release
instruction to maintain the sequential order among ordering instructions (RC5
listed in Section~\ref{sec:rc_rules}); the \vr register is not
updated. Since \vr does not increase, we assign the same version to post-fence
memory accesses thereby not placing any additional restrictions on them.
}

\begin{verule}
\label{lem:strl_version}
If a store-release fence separates two regular memory accesses a and b, then
the versions assigned for the three memory accesses are as follows:
\begin{center}
$  a \po stlr \po b \implies v_a = v_b < v_{stlr}$
\end{center}
\end{verule}

\ignore{
\subsection{Model Equivalence}

In this section, we explain how the versioning semantics listed in the
previous section satisfy the $RC_{SC}$ semantics described in
Section~\ref{sec:rc_rules}.
}

\section{Architecture}
\label{sec:louvre}

This section details a low overhead implementation that utilizes \method to
enforce the prescribed ordering. In the current work, we assume an
architecture that implements the conventional $RC_{SC}$ memory
model~\cite{gha:len90} and employs a load-store queue and a store buffer.  We
augment these structures to hold version information for each entry.

\subsection{Hardware Structures}
\label{sec:hardware}

\begin{figure}[!h]
\includegraphics[width=0.5\textwidth]{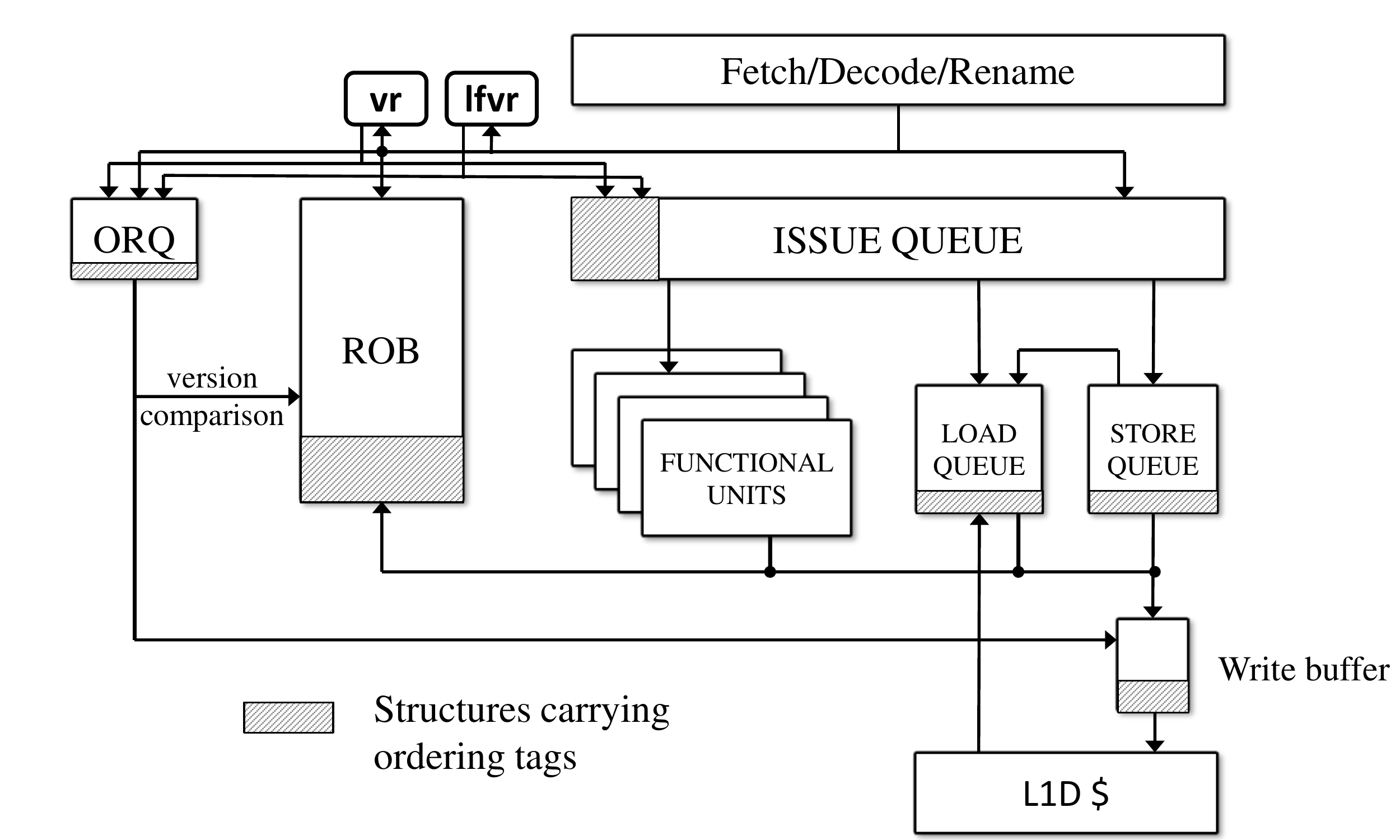}
\caption{Overview of a simplified architecture implementing \method{}.}
\label{fig:architecture_pipeline}
\end{figure}

We use the following structures to support versioning.

\noindent
\textbf{Ordering version}: A tag added to each load and store that tracks
  the assigned version.

\noindent
\textbf{Version registers} (\vr and \lfvr): We assign a version to ordinary memory
accesses during issue using the \vr register. The \lfvr register is the last
fence version register and keeps track of the version of the last issued
fence. This is used to assign versions to ordering instructions such that they
are in sequential order ($RC5$ in Section~\ref{sec:rc_rules}.

\noindent
\textbf{Min version registers} (\vsb and \vlsq): These registers keeps track of the
lowest version of the entries in the store buffer and LSQ respectively.

\noindent
\textbf{ORQ} (\orq): We need to track in-flight synchronizing loads and full
fences to squash speculative post-fence loads upon invalidation. We use an
ordering queue to track these fences.

A simplified architecture with the above structures is detailed in
Fig.~\ref{fig:architecture_pipeline}.

\ignore{
\comms{repeated information}
In a conventional implementation, we retire a fence instruction from the ROB
only after the fence enforces its ordering constraints, that is, we track the
ordering constraints of the fence instruction by its entry in the ROB. In
versioning, the assigned version encodes the ordering constraints enforced by
fence instructions. We use \vsb to detect if there are po-before stores
pending in the store buffer before retiring loads from the ROB. If the version
of the load at the head of the ROB is greater than \vsb, we cannot retire the
load since it will be visible before the store with a lower version in the
store buffer.

Upon receiving a cache invalidation request, we compare the versions of
\textit{speculative} loads that read their values from this cache line to \vsb
and \vlsq. If the version of the speculative loads is higher, they might have
read stale values possibly violating the ordering constraints. So, we squash
all such higher versioned loads along with their dependent instructions and
re-execute them.
}

\subsection{Version Registers Update}

We increment the last fence version register, \lfvr, after issuing any fence
instruction. This is done to ensure that the ordering instructions are
properly ordered.

We update the version register \vr to \lfvr when we issue a full fence
instruction. This updated version register ensures that the version to later
memory access is higher and hence orders them with previous memory accesses.

Table~\ref{tab:ordering_update} summarizes the assignment of versions and
updates to the version register \vr after issuing an instruction.  Since we
continuously increment the version registers, depending on its size, it will
occasionally overflow. To handle this overflow, we stop issuing any further
instructions and wait for the pipeline to drain. After draining the pipeline,
we reset the version registers and start reissuing instructions. Assuming that
we issue 10 fence instructions per kilo instructions, with a 10-bit version
register we expect to see an overflow approximately once every hundred
thousand instructions.

\begin{table}[h]
  \caption{Version assignment and register update on issuing an
    instruction.}
\begin{center}
  \begin{tabular}{c|c|l}
  \toprule
  Issued Inst. & Version assignment & Register update \\
  \hline
  load/store & version = \textbf{vr} & \vr is not updated \\
  \hline
  load-acquire & version = \textbf{vr} & \vr is not updated \\ 
  \hline
  store-release & version = \textbf{vr} + 1 & \vr is not updated \\
  \hline
  fence & --- & \textbf{vr} = \textbf{lfvr} \\
  \bottomrule
\end{tabular}
\label{tab:ordering_update}
\end{center}
\end{table}

We illustrate an example of version assignment to a
  sequence of memory operations in Table~\ref{tab:complete_example}.
\begin{table}
  \caption{Worked out example of versioned memory operations. Operation
    \textit{mi} can either be a load or store. v(m) is the version assigned to
    each memory access.}
\begin{center}
\begin{tabular}{lrrr}
  \toprule
Operation & v(m) & lfvr & vr \\
Initial value & - & 0 & 0\\
  \hline
m1 & 0 & 0 & 0 \\
ldar(m2) & 0 & 1 & 0\\
m3 & 0 & 1 & 0\\
stlr(m4) & 1 & 2 & 0\\
m5 & 0 & 2 & 0\\
fence & - & 3 & 3\\
m6 & 3 & 3 & 3\\
  \hline
\end{tabular}
\end{center}
\label{tab:complete_example}
\end{table}

\subsection{Retirement}
\label{sec:retirement}

In this section we detail the conditions required for the retirement of
different kinds of instructions in \method.

\noindent
\textbf{Stores}: A store reaching the head of ROB will retire
immediately. Retired stores move to the store buffer awaiting completion.

\noindent
\textbf{Loads}: In conventional implementations, a satisfied load that reaches
the head of the ROB can retire. However, in \method, we need to consider the
stores residing in the store buffer because of the relaxed constraints on
fence retirement. In the proposed mechanism, a load cannot retire as long as a
store exists in the store buffer with a lower version. This is to ensure that
any store pending in the store buffer with a lower version completes before
this load and hence is ordered before it.  Consider the case where a
$store A; fence; load B$ sequence of instructions execute. We assign version 1
to \textit{store A} and version 2 to \textit{load B}.  \textit{load B} should
not retire until \textit{store A} completes from the store buffer even if it
is at the head of ROB.

To check for a store versioned lower than the load at the head of the ROB
without probing the store buffer, we use the \vsb. As explained in
Section~\ref{sec:hardware}, the version in \vsb is the minimum version of the
stores in the store buffer. Hence, if the version of the load at the head of
the ROB is not greater than the \vsb, then we satisfy all the ordering
constraints and can retire the load at the head of the ROB.

\noindent
\textbf{Ordering Instructions} A synchronizing store, similar to a store,
retires once it reaches the head of the ROB and then moves to the store
buffer. There is no other constraint on it since the version assigned to its
store ensures the required ordering. A satisfied synchronizing load, similar
to a load, retires once it reaches the head of the ROB and its version is not
greater than the version in the \vsb. A full fence instruction retires once it
reaches the head of the ROB. There is no delay in retiring a full fence
instruction in the proposed technique, since versioning ensures ordering on
the memory accesses for which these fences are responsible.

\vspace{-0.15cm}
\subsection{Store Completion}
\label{subsec:completion}


\textit{Stores} complete from the store buffer once they update the cache
line. Since the store buffer is unordered, it allows out-of-order completion;
the stores can complete in an order according to the versions assigned. For
completion, of all the stores that have their cache line available, one of the
lowest versioned stores completes first. This ensures that the stores complete
in the order enforced by the versions. We also complete the oldest store
residing in the store buffer irrespective of its version, since no ordering
constraint can exist on the oldest store.

\subsection{Speculative Execution and Invalidation}
\label{sec:spec_exec_inv}

As discussed in Section~\ref{sec:load_spec}, we do not need to squash and
re-execute all speculative loads when we invalidate the cache line that a load
reads from. We need to squash loads only in cases where there is an in-flight
full fence or a load-acquire instruction enforcing order on it. We ensure this
order by comparing the versions of the loads that need invalidation and the
versions \vsb and \vlsq. We squash and re-execute speculative loads
\textit{only if} the version of the load is greater than these versions. Only
post-fence loads with an active fence (detected using \orq) have a version
that is higher than the version in \vsb and \vlsq. Squashing such loads
ensures that it is not ordered ahead of the pre-fence stores, which may reside
in the store buffer. In all the other cases, we do not squash loads upon a
cache invalidation. This increases the effectiveness of speculation in the
presence of ordering instructions.

\begin{figure}[!h]
\hspace{-1.8em}\includegraphics[width=0.55\textwidth]{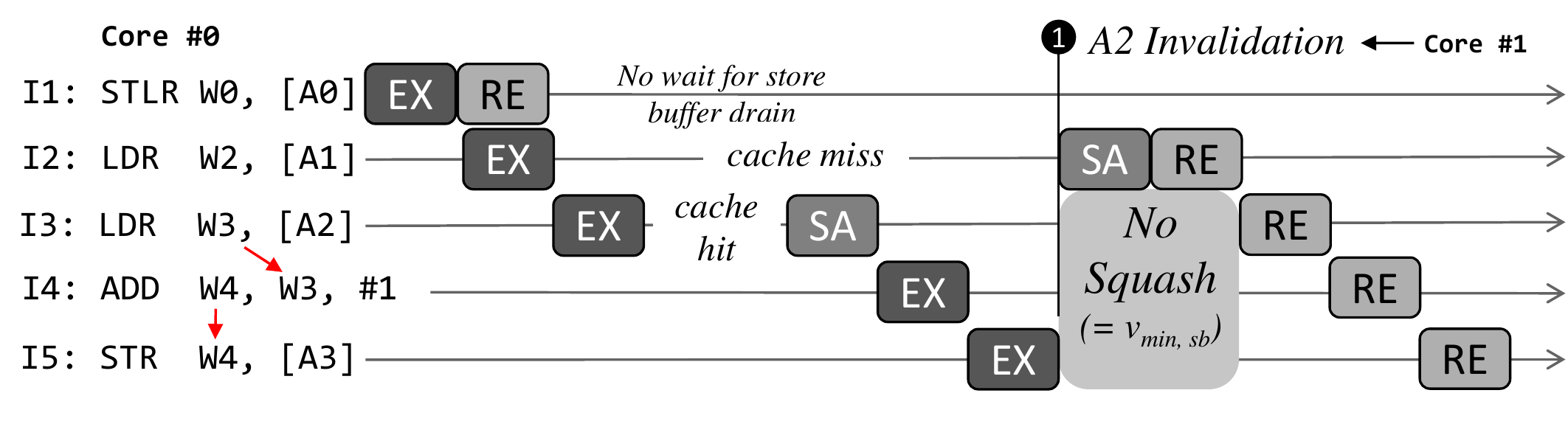}
\caption{Efficient load speculation in \method}
\label{fig:load_speculation_versioning}
\end{figure}

We illustrate an optimal execution of the scenario shown in
Figure~\ref{fig:load_speculation} when using versioning in
Figure~\ref{fig:load_speculation_versioning}. In this example, instruction
$I1$ is a store-release fence. Once it retires, the ordering store moves to
the store buffer with a version of 1. The version assigned to loads $I2$ and
$I3$ is 0. When the invalidation message for cache line containing $A2$
arrives at \mynode{1}, we can determine that no ordering constraint exists by
comparing the version of the load to the \vsb. Since the version of the load
is lower than \vsb, it allows us to skip squashing $I3$ thereby saving the
re-execution penalty.

\subsection{Versioning Example}
\label{sec:ver_example}

\begin{figure*}[!h]
\hspace{-0.5em}\includegraphics[width=1.05\textwidth]{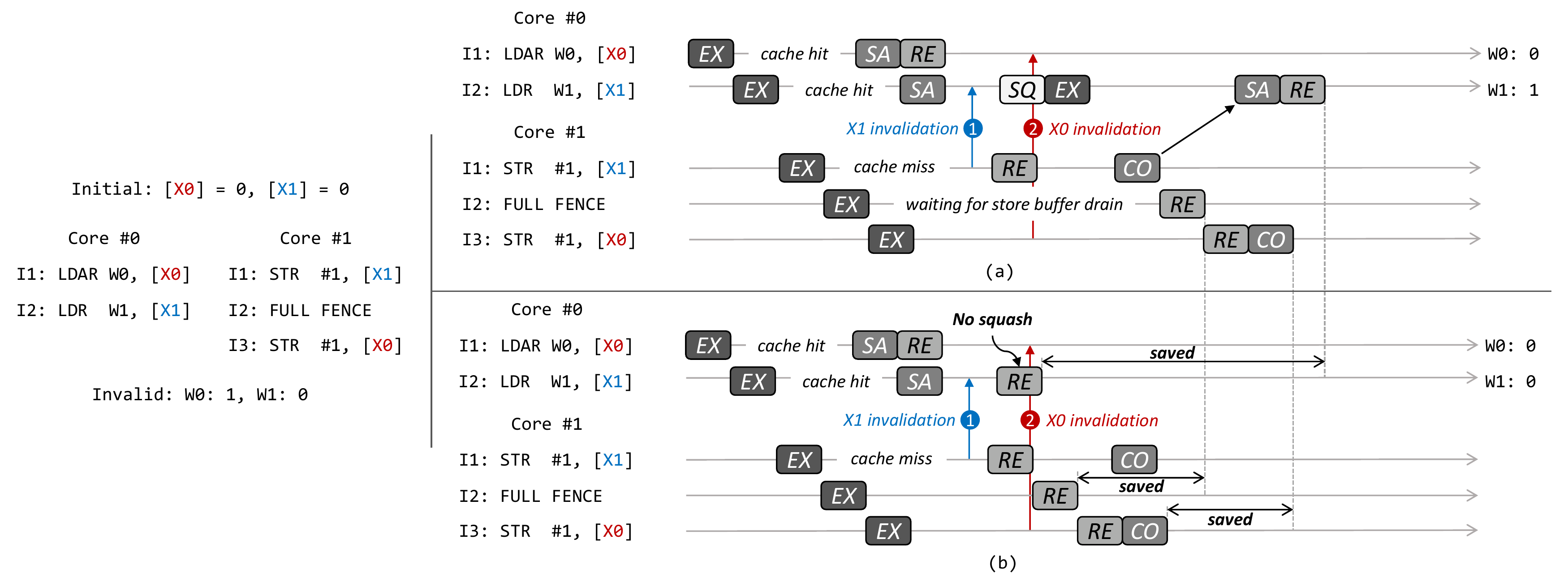}
\caption{A synchronizing load execution in (a) the base architecture
  (b) \method}
\label{fig:loadacquire_example}
\end{figure*}

We present a simple two core execution scenario for a sample instruction
sequence that has a synchronizing load in the base architecture in
Figure~\ref{fig:loadacquire_example} (a) and execution for the same instruction
sequence in \method in Figure~\ref{fig:loadacquire_example} (b).

In this example, instruction $I1$ on core 1 generates an invalidation request
when it writes to address $A1$. In the base architecture, this invalidation
request squashes the satisfied load $I2$ on core 0 since it is speculative as
shown in Figure~\ref{fig:loadacquire_example} (a). However, since the
synchronized load $I1$ on core 0 has retired by this point, squashing $I2$ is
unnecessary. In \method, we can detect this scenario by comparing the version
of the load that the processor has to squash with the minimum version
registers. Since the version of this load is not greater than the minimum
version registers, we avoid squashing the load as shown in
Figure~\ref{fig:loadacquire_example} (b). Note that both the executions
produce a valid order even though the values read are different. In the base
architecture, core 0 reads $(W0: 0, W1: 1)$ whereas in the \method
architecture, core 0 reads $(W0: 0, W1: 0)$. The only invalid combination is
$(W0: 1, W1: 0)$.

\subsection{Satisfying RC\textsubscript{SC}}

In this section, we discuss how the versioning rules described in
Section~\ref{sec:vr_update} satisfy the ordering rules of $RC_{SC}$ listed in
Section~\ref{sec:rc_rules}. We satisfy $RC1$ and $RC2$ by versioning
rule~\ref{lem:fence_version} which ensures that the assigned version is
incremented when a full fence instruction is issued. We satisfy $RC3$ by the
restriction on retirement of a synchronizing load as described in
Section~\ref{sec:ldar_fence_update}. We satisfy $RC4$ by the versioning
rule~\ref{lem:strl_version} which assigns a higher version to the
synchronizing store than the po-before memory accesses. Finally, we satisfy
$RC5$ by the squashing and retirement semantics of \method described in
Section~\ref{sec:louvre}.

\section{Analysis and Experiments}
\begin{table}
\caption{Simulated Architecture Parameters}
\begin{tabular}{ll}
  \toprule
Processor & 8-core, 2 GHz with 192 entry ROB \\
  \midrule
Fetch & 6 instructions, 2 loads or stores per cycle \\
  \midrule
Issue & Out-of-order schedule \\
  \midrule
Store Buffer & 16 and 32 entry unordered buffer \\
  \midrule
L1 Cache & 64 KB I-cache and D-cache per core \\
 & 2 cycle access latency \\
  \midrule
L2 Cache & 512 KB shared L2 cache (4 core) (LLC) \\
 & 10 cycle access latency \\
  \midrule
L3 Cache & 8 MB shared L3 cache (LLC) \\
 & 25 cycle access latency \\
  \midrule
Coherence & MESI protocol \\
  \midrule
Memory & 4 GB, 100 cycle access latency \\
  \bottomrule
\end{tabular}
\label{tab:params}
\end{table}

\ignore{
\begin{figure}
\includegraphics[width=0.45\textwidth]{figures/architecture_new.pdf}
\caption{Proposed versioned LSQ and SB architecture.}
\label{fig:architecture}
\end{figure}
}

We implement and analyze \method using 10-bit version tags. Each entry in the
LSQ and the store buffer is augmented with these tags. For a 16 entry store
buffer and a 64-entry LSQ the storage cost of these tags is (16 + 64) * 10 =
800 bits or 100 bytes. 

To implement the tracking of lowest versions in the store buffer, we use a
hierarchical comparator network~\cite{koc:tor11}. This network for a store
buffer of 16 entry size uses 15 comparators. We also use two additional
comparators to compare versions in these version registers.

In our experiments we assume a release consistency architecture loosely
modeled on an in-production ARM64 system in Apple's Cyclone
processor\cite{cyclone}. ARM64 follows the $RC_{sc}$\cite{gha:len90,armv8}
consistency model where the ordering operations are sequentially consistent.
We consider a 192 entry reorder buffer which is interfaced to the data cache
through the load-store queue. We assume an out-of-order execution core which
can speculatively execute loads and stores. Other micro-architectural details
of the simulation are listed in Table~\ref{tab:params}.  We assume an
invalidation based coherence protocol and use MESI protocol in our
experiments. So as to not complicate our simulation, we assume that all the
caches participate in the same coherence domain, i.e., any updates in one
cache will be propagated to all processors. Please note that this assumption
is stricter than that in the ARM memory model which allows non-multi-copy
atomic behavior. We assume that stores are multi-copy atomic in our model. We
evaluate our implementation on architectures using an unordered 16 entry store
buffer.

We model our simulator after the one implemented in the O3 cpu model of the
Gem5 simulator~\cite{gem5} as our baseline architecture.  In this model,
speculatively executed loads are squashed immediately upon invalidation and
all the ordering instructions disable speculative execution of post-fence
loads and stores.

We evaluate our implementations in a cycle-accurate simulator implemented
using the Qsim~\cite{ker:rod12} framework which includes SST~\cite{sst} and
MacSim~\cite{macsim}. We ran parallel benchmarks from two suites,
PARSEC~\cite{parsec} and SPLASH-2~\cite{splash}.

\section{Results}

In this section we present our evaluation of the proposed versionining
mechanism. We compare our results against the base line version with an equal
store buffer capacity. The various characteristics which we focus on to show
the efficiency of the proposed technique are as follows:

\begin{enumerate}
\item Performance evaluation (IPC).
\item Reduction in stalls caused by an ordering instruction. 
\item Reduction in store latency.
\item The average residency of a fence in ROB.
\end{enumerate}

\subsection{Performance Evaluation}

\begin{figure*}[!ht]
\includegraphics[width=0.95\textwidth]{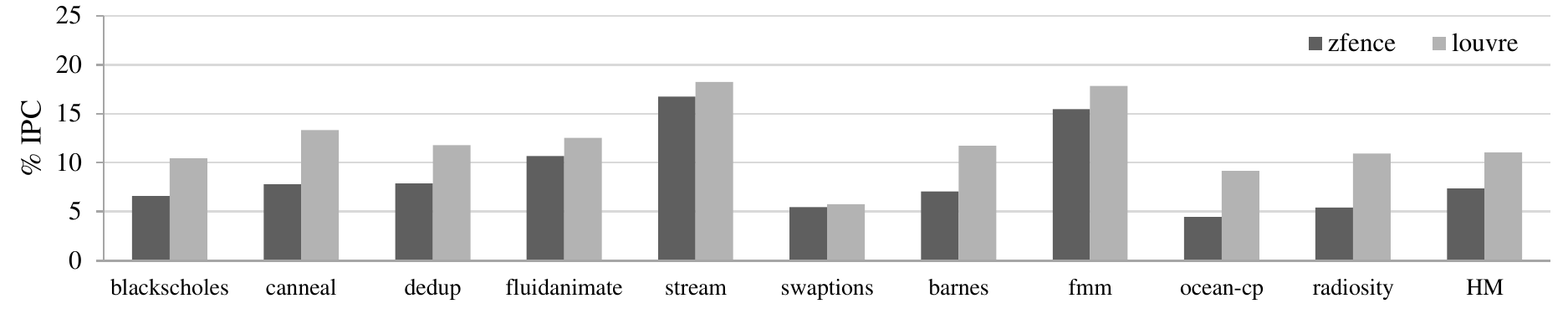}
\caption{Performance of \method when compared to \textit{zFence} for a
  16-entry store buffer.}
\label{fig:ipc}
\end{figure*}

The main performance benefit of the proposed mechanism is being able to retire
a fence instructions without waiting for the store buffer to drain. Also
speculatively executing loads and stores past the ordering instruction without
incurring any delay allows additional benefits. We keep track of the ordering
constraint using versions and the version registers when using versioning
which enables us to retire the fences ahead of time when compared to a naive
implementation. This retirement reduces the stalls in the pipeline. Also note
that since this technique is primarily implemented within the core, based on
contention for shared lines, the performance should increase with increasing
number of cores.

Figure \ref{fig:ipc} shows the performance of our versioning scheme when
compared to a recently proposed fence implementation
\textit{zFence}\cite{aga:sin15}. In the \textit{zFence} mechanism, the stores
with LLC misses request exclusive permission, which when granted allows
execution past the fence. This mechanism, similar to ours, tries to reduce the
delay of fence caused by draining the store buffer. However, there is still a
significant delay in acknowledging this permission request by the directory
controller since you need to get the permission for all the stores residing in
the store buffer before proceeding with further execution. Whereas in \method,
no such permission request needs to be issued.  On average we improve
performance by 11\% with \textit{streamcluster} gaining the most at 18\% IPC
over the baseline. When compared to \textit{zFence} we improve the performance
by 4\% on average. Although we did not perform any power consumption studies,
we believe that the performance gains will translate to energy savings.

\subsection{Scheduling Stalls}

\begin{figure}[ht]
\hspace{-0.8em}\includegraphics[width=0.5\textwidth]{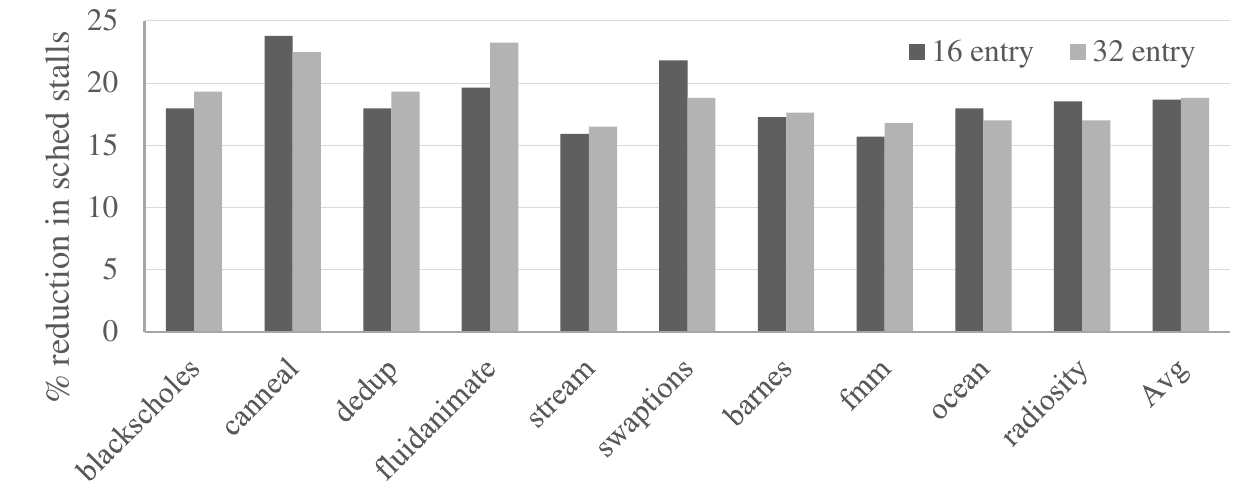}
\caption{Reduction in scheduling stalls in \method.}
\label{fig:sched_stall}
\end{figure}

As we explained in Section \ref{sec:overhead}, an ordering instruction
introduces bubbles in the pipeline causing a stall. One of the major component
of this stall is the wait for the store buffer to drain. Using a versioned
store buffer eliminates this stall. The reduction in the stalls when compared
to a conventional store buffer is shown in Figure~\ref{fig:sched_stall}. We
see an average reduction of 19\% in these stalls.

\subsection{Store latency}

\begin{figure}[ht]
\hspace{-0.8em}\includegraphics[width=0.52\textwidth]{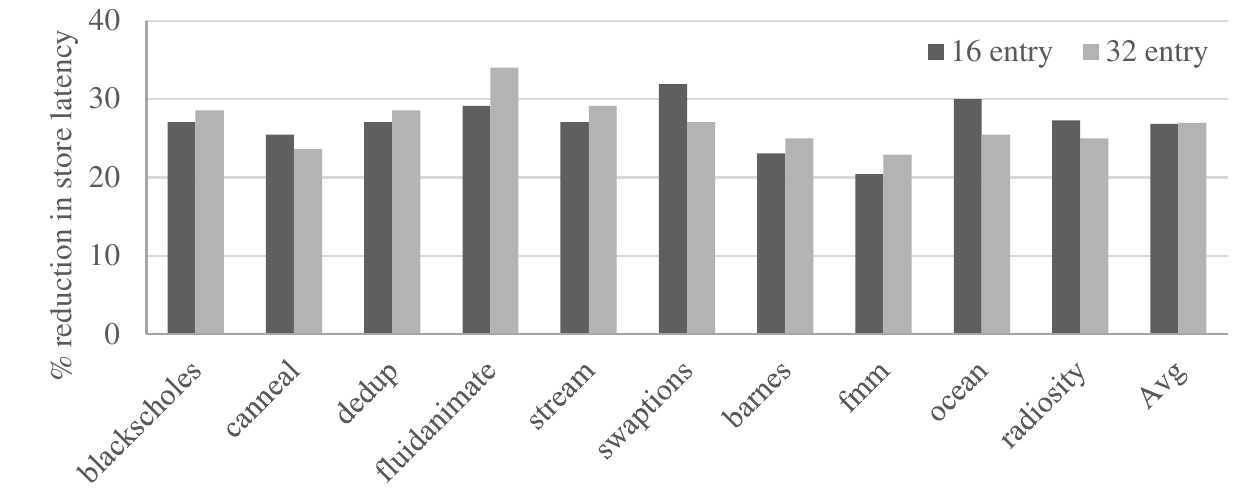}
\caption{Reduced store latency caused by early retirement}
\label{fig:store_latency}
\end{figure}

Scheduling stores without any fence induced delay allows us to issue a memory
request earlier for it. Also, since the stores with the same version can
complete out of order from the store buffer, the store latency is reduced from
an 63 cycles to 46 cycles on average, which is a ~26\% reduction as shown in
Figure \ref{fig:store_latency}. It should be noted that savings in store
latency do not directly transfer to increase in performance. This is because
of other architectural optimizations implemented that hide memory access
latency like prefetching and speculative execution even in the baseline
architecture.

\subsection{Early Execution}

\begin{figure}
\hspace{-0.8em}\includegraphics[width=0.5\textwidth]{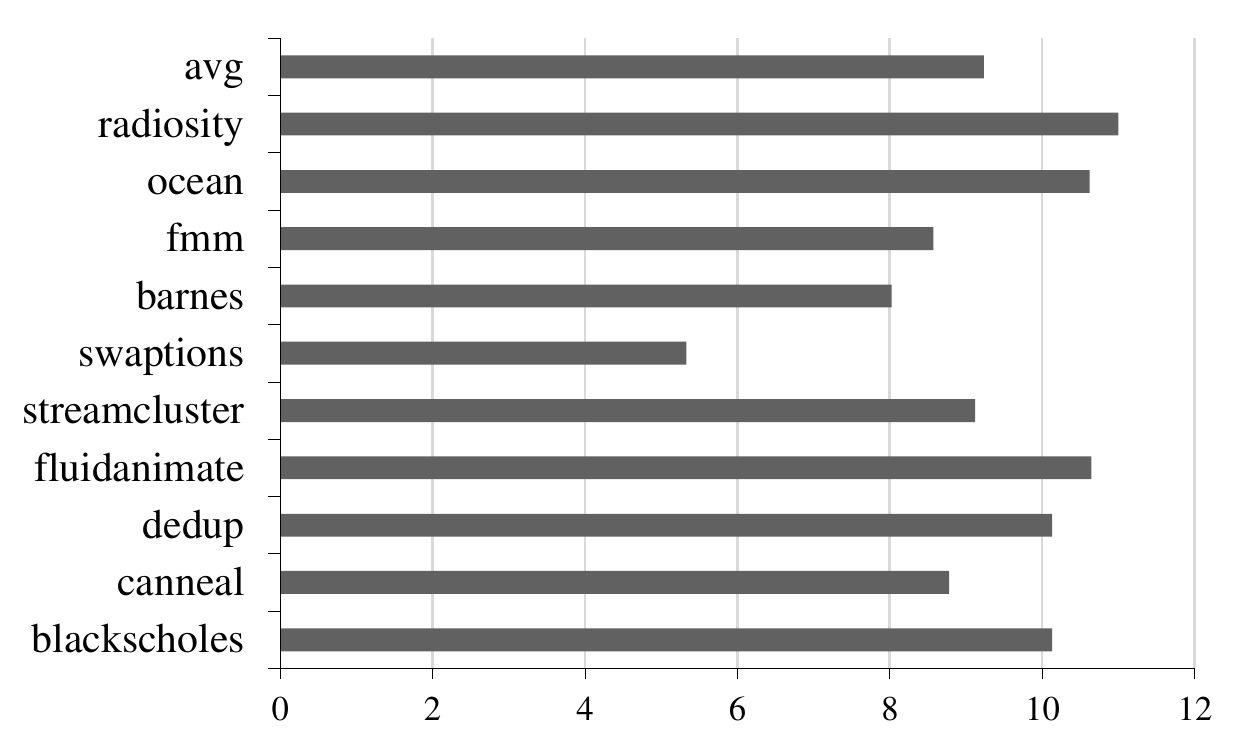}
\caption{Average reduction in stalled cycles for post-fence memory accesses
  when using a 16-entry versioned store buffer.}
\label{fig:early_execution}
\end{figure}

Whenever a full fence or an acquire fence is active, we do not speculatively
execute post-fence memory accesses as this can violate the memory consistency
guarantees. Using a versioned store buffer allows us to relax this constraint,
thereby, enabling us to speculatively execute post-fence memory accesses. The
reduction in stalled cycles in camparison to a naive store buffer
implementation is shown in Fig. \ref{fig:early_execution}.  On average we find
that a post-fence store in a versioned store buffer implementation starts
executing ~8 cycles early (contributing to ~9\% reduction in latency) when
compared to an implementation with naive store buffer.

\subsection{Fence Residency in ROB}

\begin{figure}[!h]
\hspace{-0.8em}\includegraphics[width=0.5\textwidth]{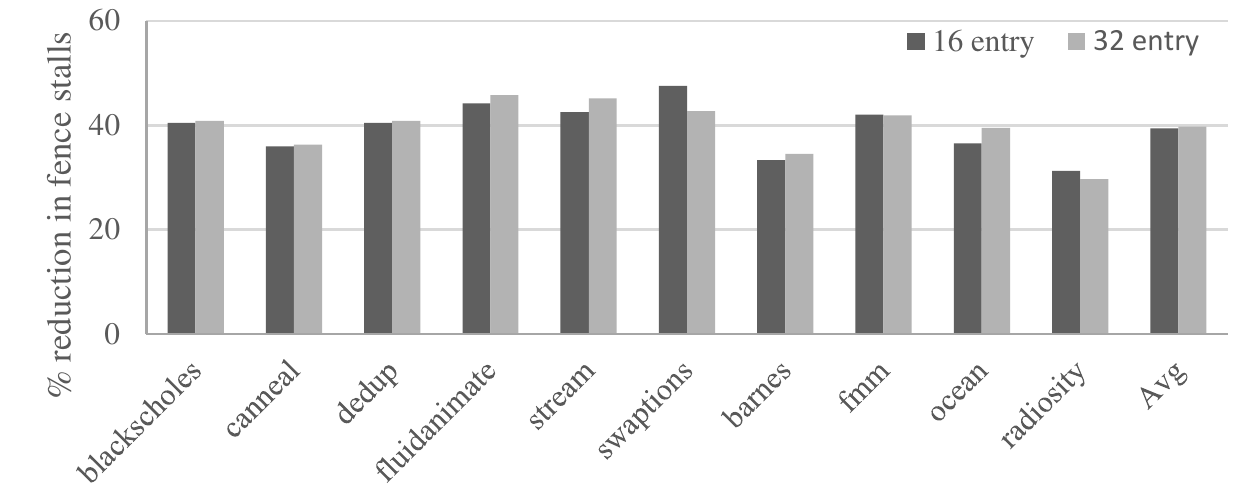}
\caption{Reduced residency of fences in \method.}
\label{fig:fence_stall}
\end{figure}

An instruction is allocated an entry in the ROB once it is scheduled. This
entry remains until the instruction is retired from the ROB. A fence
instruction retires once it enforces all its consistency guarantees.  A naive
store buffer needs to be drained completely before a fence can be retired and
removed from the ROB. A versioned store buffer, on the other hand allows us to
retire the fence without any delays (as explained in
Section~\ref{sec:retirement}) from the ROB as the version assigned to the
memory access will be responsible for ensuring ordering. This reduces the
lifetime of a fence in the ROB. In other words, it reduces the number of
cycles a fence resides in the ROB. This fence residency goes down from an
average of 142 cycles in the base implementation to 82 cycles in the versioned
implementation, a reduction of 39.6\% as shown in Fig. \ref{fig:fence_stall}.

\subsection{Congested Store Buffer}

\begin{figure}[!h]
\includegraphics[width=0.5\textwidth]{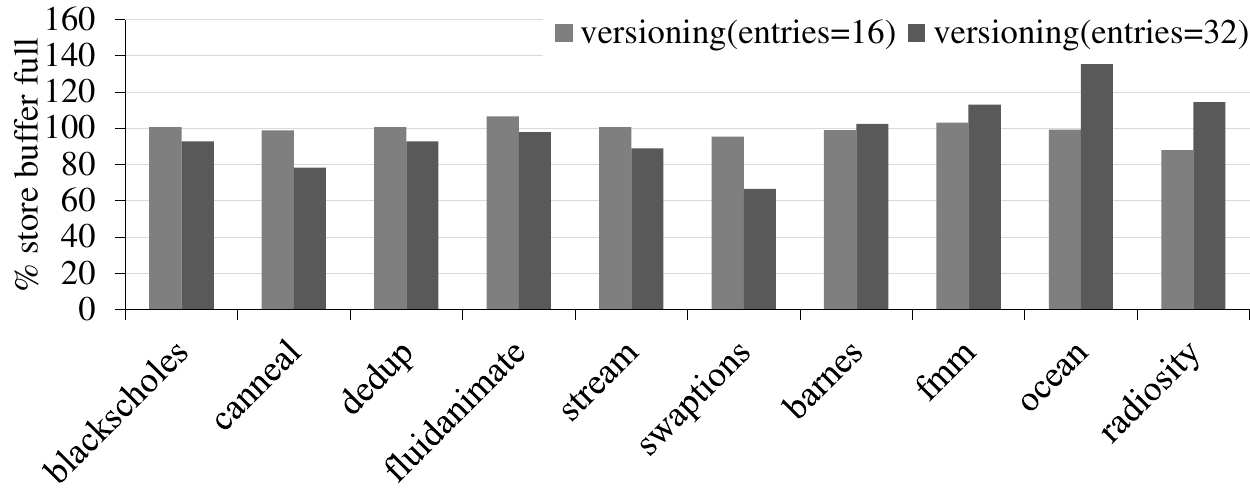}
\caption{Stalls caused by the store buffer being full}
\label{fig:wb_full}
\end{figure}

One side-effect of delaying scheduling of stores past an ordering instruction
is not being able to issue a memory request for them. Pre-fetching can
mitigate this issue, though you get it as a side-effect with \method.  As a
result of not issuing a memory request early, there might be multiple
post-fence stores with pending memory requests. When these reach the head of
the ROB, they are retired and placed in the store buffer. This causes
congestion and leads to further stalls. Figure \ref{fig:wb_full} shows the
reduction in such stalls when we employ versioning. In all but two benchmarks,
we have a decrease in the number of stalls caused by a congested write
buffer. In the two benchmarks \textit{ocean} and \textit{radiosity} using a
32-entry store buffer increases the number of stalls. This is because these
benchmarks are store intensive, and we issue lots of memory requests causing
greater cache pressure.

\section{Discussion}
\label{discussion}

\subsection{Multi-copy Atomicity}

Architectures like ARM and POWER are not multi-copy atomic~\cite{armv8_arm,
  flu:gra16}. This property allows a store to become visible to only to some
cores and not others. One of the reasons for this behavior is the sharing of
cache by some cores. An update from one core can become visible to the cores
sharing cache with it before the update propagates to the memory. Intuitively,
the cache coherence protocol should ensure that a reader sees the latest
update in the entire cache hierarchy. However, if we restrict the coherence
domain to the caches of a core cluster, it will give rise to this behavior.
Another reason for this behavior is when multiple hardware threads share the
same core using simultaneous multi-threading (SMT) as seen in POWER
processors. In such processors, the hardware threads share the store buffer
from where they can read the value of an earlier write from another hardware
thread before others. However, in our modeling, to keep our simulation
simple, we assume that all the caches participate in the same coherence domain
and that there are no shared store buffers. We plan to explore the
non-multi-copy atomic behavior in our future work.

\vspace{-1ex}
\subsection{Branch Prediction}

Since we introduce two version registers \vr and \lfvr that are updated on
instruction issue, we need to consider the scenario when branch mis-prediction
occurs. We need to make sure that the version registers have correct values
when the processor starts fetching instructions from the correct path after
recovering from a mis-prediction. We ensure this by check-pointing
registers along with the process state on speculation. On mis-predicting, we
flush the pipeline and restore the version registers from the check-pointed
state.

\subsection{Load-Store Queue Design}
\label{sec:lsq}

We discuss two possible implementation choices for the LSQ: the snoopy and the
insulated queue~\cite{cai:lip04}.  In a snoopy queue implementation, when a
cache line is invalidated, the LSQ is probed to squash all speculative loads
that were satisfied from the invalidated cache line to prevent ordering
violations. In an insulated queue implementation, this violation is prevented
at the time of retiring a load. During retirement, the LSQ conservatively
squashes all later loads in program order satisfied ahead of the current load
that can potentially violate consistency. Whereas processors with a stricter
consistency model usually employ a snoopy queue design, those with a weaker
consistency model use an insulated queue design. After all, in a strict
consistency model, we need to enforce ordering on a large number of in-flight
loads and stores. Since ordering violations in a weak consistency processor
are infrequent, an insulated load queue is usually utilized at the cost of
conservative performance. However, since a fence instruction enforces
consistency, an insulated queue implementation can be used to avoid
unnecessary squashes. In such a case, at the time of load retirement, we check
for an active fence instruction, and if such a fence is found, we identify all
the satisfied loads that are later in program order and squash them.

\subsection{Write Combining}

A write combining buffer merges multiple pending writes to the same address to
reduce the number of memory transactions. However, in \method, we ensure that
the buffer only merges stores with the same version. If it combines stores
with different versions, then an ordering violation can occur.

\vspace{-1ex}
\section{Related Work}

Considerable previous work has been done with a focus on reducing the cost of
fences in both software and hardware. The main impetus for research in this
area is that a considerable fraction of fences are determined to be
unnecessary at run-time, or that they are ensuring ordering on memory accesses
for which ordering is unnecessary, for example, thread-local memory
accesses. Various hardware and software techniques were proposed to remove the
fence overhead.  Software techniques focus mainly on compiler techniques which
automatically infer fences~\cite{kup:vec10}, identify and remove redundant
fences~\cite{vaf:nar11} or which reduce the strength of a fence i.e., convert
a stronger fence to a weaker fence. Hardware techniques focused on utilizing
the dynamic information available at the time of execution of the fence to
reduce or elide the overhead of ensuring ordering. Ordering needs to be
ensured only for shared memory locations since \textit{sequentially
  consistent} (SC) semantics are guaranteed for local accesses i.e., a
processor always has a sequentially consistent view of its own memory
accesses. Taking the help of the compiler to determine such memory accesses
and using an un-ordered store buffer for the identified accesses reduces the
overhead of fence instruction~\cite{sin:nar12}.

Lin et al. proposed address-aware fences~\cite{lin:nag14} in which the shared
memory addresses on which ordering is to be ensured are tracked in a watch
list. Ordering is only ensured for these memory locations. All other memory
accesses are unordered. The authors extend the idea to order memory accesses
only within the scope of the fence~\cite{lin:nag14}. Singh et al. in
~\cite{sin:nar12} proposed an approach to design sequentially consistent
hardware in which they classify memory accesses as either local or shared and
try to maintain order only for the shared accesses. They do this by using an
unordered store buffer for local memory accesses in addition to a regular FIFO
store buffer for shared memory accesses. The local memory accesses are
processed through the unordered store buffer avoiding the ordering overhead on
them.

\section{Conclusion}

A processor implementing a weak memory model requires higher number of fence
instructions to ensure memory consistency for parallel programs than a
processor implementing stronger memory models. However, the weaker semantics
provide new opportunities for extracting performance by allowing us more
leeway to reorder memory operations. Uni-directional fences allow us to only
restrict certain possible reordering ensuring release consistency. In this
paper, we propose a light-weight mechanism which assigns versions to in-flight
memory accesses. We detailed how fences can be retired without causing any
significant delays utilizing versions. With minor modifications to the
micro-architecture where we add version holding structures, we significantly
reduce the fence overhead. We also studied the various architectural
optimizations that can be implemented in terms of memory re-orderings and
speculative execution of memory accesses when uni-directional fence
instructions are employed and presented a design for such an optimal
implementation. We found that implementing \method reduces pipeline stalls
caused by fences and store latency thereby increasing performance by 11\% on
average over the baseline and 4\% over \textit{zFence}.






\bibliographystyle{IEEEtran}
\bibliography{IEEEabrv,ref}

\begin{thebibliography}{10}
\providecommand{\url}[1]{#1}
\csname url@samestyle\endcsname
\providecommand{\newblock}{\relax}
\providecommand{\bibinfo}[2]{#2}
\providecommand{\BIBentrySTDinterwordspacing}{\spaceskip=0pt\relax}
\providecommand{\BIBentryALTinterwordstretchfactor}{4}
\providecommand{\BIBentryALTinterwordspacing}{\spaceskip=\fontdimen2\font plus
\BIBentryALTinterwordstretchfactor\fontdimen3\font minus
  \fontdimen4\font\relax}
\providecommand{\BIBforeignlanguage}[2]{{%
\expandafter\ifx\csname l@#1\endcsname\relax
\typeout{** WARNING: IEEEtran.bst: No hyphenation pattern has been}%
\typeout{** loaded for the language `#1'. Using the pattern for}%
\typeout{** the default language instead.}%
\else
\language=\csname l@#1\endcsname
\fi
#2}}
\providecommand{\BIBdecl}{\relax}
\BIBdecl

\bibitem{blu:joe96}
R.~D. Blumofe, C.~F. Joerg, B.~C. Kuszmaul, C.~E. Leiserson, K.~H. Randall, and
  Y.~Zhou, ``Cilk: an efficient multithreaded runtime system,'' in \emph{PPOPP
  '95: Proc. of the fifth ACM SIGPLAN Symp. on Principles and practice of
  parallel programming}, 1995.

\bibitem{dor:rod05}
A.~J. Dorta, C.~Rodr\'{\i}guez, F.~de~Sande, and A.~Gonzalez-Escribano, ``The
  {OpenMP Source Code Repository},'' \emph{Parallel, Distributed, and
  Network-Based Processing, Euromicro Conf. on}, 2005.

\bibitem{adve:hill90}
\BIBentryALTinterwordspacing
S.~V. Adve and M.~D. Hill, ``Weak ordering --- a new definition,'' in
  \emph{Proceedings of the 17th Annual International Symposium on Computer
  Architecture}, ser. ISCA '90.\hskip 1em plus 0.5em minus 0.4em\relax New
  York, NY, USA: ACM, 1990, pp. 2--14. [Online]. Available:
  \url{http://doi.acm.org/10.1145/325164.325100}
\BIBentrySTDinterwordspacing

\bibitem{sev11}
\BIBentryALTinterwordspacing
J.~\v{S}ev\v{c}\'{\i}k, ``Safe optimisations for shared-memory concurrent
  programs,'' in \emph{Proceedings of the 32Nd ACM SIGPLAN Conference on
  Programming Language Design and Implementation}, ser. PLDI '11.\hskip 1em
  plus 0.5em minus 0.4em\relax New York, NY, USA: ACM, 2011, pp. 306--316.
  [Online]. Available: \url{http://doi.acm.org/10.1145/1993498.1993534}
\BIBentrySTDinterwordspacing

\bibitem{eid:reg08}
\BIBentryALTinterwordspacing
E.~Eide and J.~Regehr, ``Volatiles are miscompiled, and what to do about it,''
  in \emph{Proceedings of the 8th ACM International Conference on Embedded
  Software}, ser. EMSOFT '08.\hskip 1em plus 0.5em minus 0.4em\relax New York,
  NY, USA: ACM, 2008, pp. 255--264. [Online]. Available:
  \url{http://doi.acm.org/10.1145/1450058.1450093}
\BIBentrySTDinterwordspacing

\bibitem{tra:pra06}
\BIBentryALTinterwordspacing
O.~Trachsel, C.~von Praun, and T.~R. Gross, ``On the effectiveness of
  speculative and selective memory fences,'' in \emph{Proceedings of the 20th
  International Conference on Parallel and Distributed Processing}, ser.
  IPDPS'06.\hskip 1em plus 0.5em minus 0.4em\relax Washington, DC, USA: IEEE
  Computer Society, 2006, pp. 33--33. [Online]. Available:
  \url{http://dl.acm.org/citation.cfm?id=1898953.1898968}
\BIBentrySTDinterwordspacing

\bibitem{cho:fah04}
\BIBentryALTinterwordspacing
Y.~Chou, B.~Fahs, and S.~Abraham, ``Microarchitecture optimizations for
  exploiting memory-level parallelism,'' \emph{SIGARCH Comput. Archit. News},
  vol.~32, no.~2, pp. 76--, Mar. 2004. [Online]. Available:
  \url{http://doi.acm.org/10.1145/1028176.1006708}
\BIBentrySTDinterwordspacing

\bibitem{leb:kop02}
A.~R. Lebeck, J.~Koppanalil, T.~Li, J.~Patwardhan, and E.~Rotenberg, ``A large,
  fast instruction window for tolerating cache misses,'' in \emph{ISCA-29},
  2002, pp. 59--70.

\bibitem{luk01}
C.-K. Luk, ``Tolerating memory latency through software-{C}ontrolled
  pre-execution in simultaneous multithreading processors,'' in
  \emph{ISCA-23}.\hskip 1em plus 0.5em minus 0.4em\relax New York, NY, USA:
  ACM, 2001, pp. 40--51.

\bibitem{mar:tor02}
\BIBentryALTinterwordspacing
J.~F. Mart\'{\i}nez and J.~Torrellas, ``Speculative synchronization: Applying
  thread-level speculation to explicitly parallel applications,'' \emph{SIGOPS
  Oper. Syst. Rev.}, vol.~36, no.~5, pp. 18--29, Oct. 2002. [Online].
  Available: \url{http://doi.acm.org/10.1145/635508.605400}
\BIBentrySTDinterwordspacing

\bibitem{ran:pai97}
\BIBentryALTinterwordspacing
P.~Ranganathan, V.~S. Pai, and S.~V. Adve, ``Using speculative retirement and
  larger instruction windows to narrow the performance gap between memory
  consistency models,'' in \emph{Proceedings of the Ninth Annual ACM Symposium
  on Parallel Algorithms and Architectures}, ser. SPAA '97.\hskip 1em plus
  0.5em minus 0.4em\relax New York, NY, USA: ACM, 1997, pp. 199--210. [Online].
  Available: \url{http://doi.acm.org/10.1145/258492.258512}
\BIBentrySTDinterwordspacing

\bibitem{gni:fal99}
\BIBentryALTinterwordspacing
C.~Gniady, B.~Falsafi, and T.~N. Vijaykumar, ``Is sc + ilp = rc?''
  \emph{SIGARCH Comput. Archit. News}, vol.~27, no.~2, pp. 162--171, May 1999.
  [Online]. Available: \url{http://doi.acm.org/10.1145/307338.300993}
\BIBentrySTDinterwordspacing

\bibitem{dua:muz13}
\BIBentryALTinterwordspacing
Y.~Duan, A.~Muzahid, and J.~Torrellas, ``Weefence: Toward making fences free in
  tso,'' \emph{SIGARCH Comput. Archit. News}, vol.~41, no.~3, pp. 213--224,
  Jun. 2013. [Online]. Available:
  \url{http://doi.acm.org/10.1145/2508148.2485941}
\BIBentrySTDinterwordspacing

\bibitem{dua:fen09}
\BIBentryALTinterwordspacing
Y.~Duan, X.~Feng, L.~Wang, C.~Zhang, and P.-C. Yew, ``Detecting and eliminating
  potential violations of sequential consistency for concurrent c/c++
  programs,'' in \emph{Proceedings of the 7th Annual IEEE/ACM International
  Symposium on Code Generation and Optimization}, ser. CGO '09.\hskip 1em plus
  0.5em minus 0.4em\relax Washington, DC, USA: IEEE Computer Society, 2009, pp.
  25--34. [Online]. Available: \url{http://dx.doi.org/10.1109/CGO.2009.29}
\BIBentrySTDinterwordspacing

\bibitem{cez:tuc07}
\BIBentryALTinterwordspacing
L.~Ceze, J.~Tuck, P.~Montesinos, and J.~Torrellas, ``Bulksc: Bulk enforcement
  of sequential consistency,'' \emph{SIGARCH Comput. Archit. News}, vol.~35,
  no.~2, pp. 278--289, Jun. 2007. [Online]. Available:
  \url{http://doi.acm.org/10.1145/1273440.1250697}
\BIBentrySTDinterwordspacing

\bibitem{lee:sim15}
\BIBentryALTinterwordspacing
J.~H. Lee, J.~Sim, and H.~Kim, ``Bssync: Processing near memory for machine
  learning workloads with bounded staleness consistency models,'' in
  \emph{Proceedings of the 2015 International Conference on Parallel
  Architecture and Compilation (PACT)}, ser. PACT '15.\hskip 1em plus 0.5em
  minus 0.4em\relax Washington, DC, USA: IEEE Computer Society, 2015, pp.
  241--252. [Online]. Available: \url{http://dx.doi.org/10.1109/PACT.2015.42}
\BIBentrySTDinterwordspacing

\bibitem{vor:kod14}
\BIBentryALTinterwordspacing
K.~Vora, S.~C. Koduru, and R.~Gupta, ``Aspire: Exploiting asynchronous
  parallelism in iterative algorithms using a relaxed consistency based dsm,''
  \emph{SIGPLAN Not.}, vol.~49, no.~10, pp. 861--878, Oct. 2014. [Online].
  Available: \url{http://doi.acm.org/10.1145/2714064.2660227}
\BIBentrySTDinterwordspacing

\bibitem{lin:nag13}
\BIBentryALTinterwordspacing
C.~Lin, V.~Nagarajan, and R.~Gupta, ``Address-aware fences,'' in
  \emph{Proceedings of the 27th International ACM Conference on International
  Conference on Supercomputing}, ser. ICS '13.\hskip 1em plus 0.5em minus
  0.4em\relax New York, NY, USA: ACM, 2013, pp. 313--324. [Online]. Available:
  \url{http://doi.acm.org/10.1145/2464996.2465015}
\BIBentrySTDinterwordspacing

\bibitem{dua:hon15}
Y.~Duan, N.~Honarmand, and J.~Torrellas, ``Asymmetric memory fences: Optimizing
  both performance and implementability,'' in \emph{Proceedings of the
  Twentieth International Conference on Architectural Support for Programming
  Languages and Operating Systems}.\hskip 1em plus 0.5em minus 0.4em\relax ACM,
  2015, pp. 531--543.

\bibitem{gha:len90}
\BIBentryALTinterwordspacing
K.~Gharachorloo, D.~Lenoski, J.~Laudon, P.~Gibbons, A.~Gupta, and J.~Hennessy,
  ``Memory consistency and event ordering in scalable shared-memory
  multiprocessors,'' in \emph{Proceedings of the 17th Annual International
  Symposium on Computer Architecture}, ser. ISCA '90.\hskip 1em plus 0.5em
  minus 0.4em\relax New York, NY, USA: ACM, 1990, pp. 15--26. [Online].
  Available: \url{http://doi.acm.org/10.1145/325164.325102}
\BIBentrySTDinterwordspacing

\bibitem{wat:lee16}
\BIBentryALTinterwordspacing
A.~Waterman, Y.~Lee, D.~A. Patterson, and K.~Asanović, ``The risc-v
  instruction set manual, volume i: User-level isa, version 2.1,'' EECS
  Department, University of California, Berkeley, Tech. Rep. UCB/EECS-2016-118,
  May 2016. [Online]. Available:
  \url{http://www2.eecs.berkeley.edu/Pubs/TechRpts/2016/EECS-2016-118.html}
\BIBentrySTDinterwordspacing

\bibitem{aga:sin15}
\BIBentryALTinterwordspacing
S.~Aga, A.~Singh, and S.~Narayanasamy, ``zfence: Data-less coherence for
  efficient fences,'' in \emph{Proceedings of the 29th ACM on International
  Conference on Supercomputing}, ser. ICS '15.\hskip 1em plus 0.5em minus
  0.4em\relax New York, NY, USA: ACM, 2015, pp. 295--305. [Online]. Available:
  \url{http://doi.acm.org/10.1145/2751205.2751211}
\BIBentrySTDinterwordspacing

\bibitem{flu:gra16}
\BIBentryALTinterwordspacing
S.~Flur, K.~E. Gray, C.~Pulte, S.~Sarkar, A.~Sezgin, L.~Maranget, W.~Deacon,
  and P.~Sewell, ``Modelling the armv8 architecture, operationally: Concurrency
  and isa,'' in \emph{Proceedings of the 43rd Annual ACM SIGPLAN-SIGACT
  Symposium on Principles of Programming Languages}, ser. POPL '16.\hskip 1em
  plus 0.5em minus 0.4em\relax New York, NY, USA: ACM, 2016, pp. 608--621.
  [Online]. Available: \url{http://doi.acm.org/10.1145/2837614.2837615}
\BIBentrySTDinterwordspacing

\bibitem{parsec}
C.~Bienia, S.~Kumar, J.~P. Singh, and K.~Li, ``{The {PARSEC} Benchmark Suite:
  Characterization and Architectural Implications},'' Princeton University,
  Tech. Rep. TR-811-08, 2008.

\bibitem{cho:spr05}
\BIBentryALTinterwordspacing
Y.~Chou, L.~Spracklen, and S.~G. Abraham, ``Store memory-level parallelism
  optimizations for commercial applications,'' in \emph{Proceedings of the 38th
  Annual IEEE/ACM International Symposium on Microarchitecture}, ser. MICRO
  38.\hskip 1em plus 0.5em minus 0.4em\relax Washington, DC, USA: IEEE Computer
  Society, 2005, pp. 183--196. [Online]. Available:
  \url{http://dx.doi.org/10.1109/MICRO.2005.31}
\BIBentrySTDinterwordspacing

\bibitem{wen:ail07}
\BIBentryALTinterwordspacing
T.~F. Wenisch, A.~Ailamaki, B.~Falsafi, and A.~Moshovos, ``Mechanisms for
  store-wait-free multiprocessors,'' in \emph{Proceedings of the 34th Annual
  International Symposium on Computer Architecture}, ser. ISCA '07.\hskip 1em
  plus 0.5em minus 0.4em\relax New York, NY, USA: ACM, 2007, pp. 266--277.
  [Online]. Available: \url{http://doi.acm.org/10.1145/1250662.1250696}
\BIBentrySTDinterwordspacing

\bibitem{bha:joh00}
R.~Bhargava and L.~K. John, ``Issues in the design of store buffers in
  dynamically scheduled processors,'' in \emph{Performance Analysis of Systems
  and Software, 2000. ISPASS. 2000 IEEE International Symposium on}, 2000, pp.
  76--87.

\bibitem{koc:tor11}
\BIBentryALTinterwordspacing
D.~Koch and J.~Torresen, ``Fpgasort: A high performance sorting architecture
  exploiting run-time reconfiguration on fpgas for large problem sorting,'' in
  \emph{Proceedings of the 19th ACM/SIGDA International Symposium on Field
  Programmable Gate Arrays}, ser. FPGA '11.\hskip 1em plus 0.5em minus
  0.4em\relax New York, NY, USA: ACM, 2011, pp. 45--54. [Online]. Available:
  \url{http://doi.acm.org/10.1145/1950413.1950427}
\BIBentrySTDinterwordspacing

\bibitem{cyclone}
A.~L. Shimpi, ``Apple's cyclone microarchitecture detailed,''
  http://www.anandtech.com/show/7910/apples-cyclone-microarchitecture-detailed,
  2014.

\bibitem{armv8}
R.~Grisenthwaite, ``Armv8 technology preview,'' 2011.

\bibitem{gem5}
\BIBentryALTinterwordspacing
N.~Binkert, B.~Beckmann, G.~Black, S.~K. Reinhardt, A.~Saidi, A.~Basu,
  J.~Hestness, D.~R. Hower, T.~Krishna, S.~Sardashti, R.~Sen, K.~Sewell,
  M.~Shoaib, N.~Vaish, M.~D. Hill, and D.~A. Wood, ``The gem5 simulator,''
  \emph{SIGARCH Comput. Archit. News}, vol.~39, no.~2, pp. 1--7, Aug. 2011.
  [Online]. Available: \url{http://doi.acm.org/10.1145/2024716.2024718}
\BIBentrySTDinterwordspacing

\bibitem{ker:rod12}
\BIBentryALTinterwordspacing
C.~D. Kersey, A.~Rodrigues, and S.~Yalamanchili, ``A universal parallel
  front-end for execution driven microarchitecture simulation,'' in
  \emph{Proceedings of the 2012 Workshop on Rapid Simulation and Performance
  Evaluation: Methods and Tools}, ser. RAPIDO '12.\hskip 1em plus 0.5em minus
  0.4em\relax New York, NY, USA: ACM, 2012, pp. 25--32. [Online]. Available:
  \url{http://doi.acm.org/10.1145/2162131.2162135}
\BIBentrySTDinterwordspacing

\bibitem{sst}
{Sandia National Laboratories}, ``{SST},'' {http://sst.sandia.gov}, 2015.

\bibitem{macsim}
H.~Kim, J.~Lee, N.~B. Lakshminarayana, J.~Sim, J.~Lim, and T.~Pho, ``Macsim: A
  cpu-gpu heterogeneous simulation framework user guide,'' 2012.

\bibitem{splash}
\BIBentryALTinterwordspacing
S.~C. Woo, M.~Ohara, E.~Torrie, J.~P. Singh, and A.~Gupta, ``The splash-2
  programs: Characterization and methodological considerations,'' in
  \emph{Proceedings of the 22Nd Annual International Symposium on Computer
  Architecture}, ser. ISCA '95.\hskip 1em plus 0.5em minus 0.4em\relax New
  York, NY, USA: ACM, 1995, pp. 24--36. [Online]. Available:
  \url{http://doi.acm.org/10.1145/223982.223990}
\BIBentrySTDinterwordspacing

\bibitem{armv8_arm}
\BIBentryALTinterwordspacing
``Arm architecture reference manual for armv8-a architecture profile,'' ARM.
  [Online]. Available:
  \url{https://developer.arm.com/docs/ddi0487/a/arm-architecture-reference-manual-armv8-for-armv8-a-architecture-profile}
\BIBentrySTDinterwordspacing

\bibitem{cai:lip04}
H.~W. Cain and M.~H. Lipasti, ``Memory ordering: a value-based approach,'' in
  \emph{Computer Architecture, 2004. Proceedings. 31st Annual International
  Symposium on}, June 2004, pp. 90--101.

\bibitem{kup:vec10}
\BIBentryALTinterwordspacing
M.~Kuperstein, M.~Vechev, and E.~Yahav, ``Automatic inference of memory
  fences,'' in \emph{Proceedings of the 2010 Conference on Formal Methods in
  Computer-Aided Design}, ser. FMCAD '10.\hskip 1em plus 0.5em minus
  0.4em\relax Austin, TX: FMCAD Inc, 2010, pp. 111--120. [Online]. Available:
  \url{http://dl.acm.org/citation.cfm?id=1998496.1998518}
\BIBentrySTDinterwordspacing

\bibitem{vaf:nar11}
\BIBentryALTinterwordspacing
V.~Vafeiadis and F.~Z. Nardelli, ``Verifying fence elimination optimisations,''
  in \emph{Proceedings of the 18th International Conference on Static
  Analysis}, ser. SAS'11.\hskip 1em plus 0.5em minus 0.4em\relax Berlin,
  Heidelberg: Springer-Verlag, 2011, pp. 146--162. [Online]. Available:
  \url{http://dl.acm.org/citation.cfm?id=2041552.2041566}
\BIBentrySTDinterwordspacing

\bibitem{sin:nar12}
\BIBentryALTinterwordspacing
A.~Singh, S.~Narayanasamy, D.~Marino, T.~Millstein, and M.~Musuvathi,
  ``End-to-end sequential consistency,'' in \emph{Proceedings of the 39th
  Annual International Symposium on Computer Architecture}, ser. ISCA
  '12.\hskip 1em plus 0.5em minus 0.4em\relax Washington, DC, USA: IEEE
  Computer Society, 2012, pp. 524--535. [Online]. Available:
  \url{http://dl.acm.org/citation.cfm?id=2337159.2337220}
\BIBentrySTDinterwordspacing

\bibitem{lin:nag14}
\BIBentryALTinterwordspacing
C.~Lin, V.~Nagarajan, and R.~Gupta, ``Fence scoping,'' in \emph{Proceedings of
  the International Conference for High Performance Computing, Networking,
  Storage and Analysis}, ser. SC '14.\hskip 1em plus 0.5em minus 0.4em\relax
  Piscataway, NJ, USA: IEEE Press, 2014, pp. 105--116. [Online]. Available:
  \url{http://dx.doi.org/10.1109/SC.2014.14}
\BIBentrySTDinterwordspacing

\end{thebibliography}
%

\end{document}